\pdfoutput=1

\documentclass{egpubl}
\usepackage{eurovis2021}

 \STAREurovis   

\usepackage[T1]{fontenc}
\usepackage{dfadobe}  

\usepackage{cite}  
\BibtexOrBiblatex
\electronicVersion
\PrintedOrElectronic
\ifpdf 
\usepackage[pdftex]{graphicx} 
\pdfoutput=1
\else \usepackage[dvips]{graphicx} \fi

\usepackage{egweblnk}

\usepackage{color}
\definecolor{turquoise}{cmyk}{0.65,0,0.1,0.3}
\definecolor{purple}{rgb}{0.65,0,0.65}
\definecolor{dark_green}{rgb}{0, 0.5, 0}
\definecolor{red}{rgb}{0.8, 0.2, 0.2}
\definecolor{blueish}{rgb}{0.0, 0.7, 1}
\definecolor{light_gray}{rgb}{0.7, 0.7, .7}
\definecolor{pink}{rgb}{1, 0, 1}
\definecolor{dark_red}{rgb}{0.5, 0, 0}

\usepackage{supertabular}
\usepackage{multirow}

\title[Data to Physicalization: A Survey of the Physical Rendering Process]%
      {Data to Physicalization: A Survey of the Physical Rendering Process}

\author[Djavaherpour et al.]
{\parbox{\textwidth}{\centering H.~Djavaherpour$^1$, F.~Samavati$^1$, A.~Mahdavi-Amiri$^2$, F.~Yazdanbakhsh$^1$, S.~Huron$^3$, R.~Levy$^1$, Y.~Jansen$^4$, and L.~Oehlberg$^1$}\\
\parbox{\textwidth}{\centering $^1$University of Calgary, $^2$Simon Fraser University, $^3$Institut Polytechnique de Paris, CNRS. $^4$Sorbonne Université, CNRS, ISIR}
}

\usepackage{array}
\usepackage{tabularx}
\usepackage{dirtytalk}
\usepackage{color}
\usepackage{diagbox}
\usepackage[flushleft]{threeparttable} 
\usepackage{multirow}
\usepackage{graphicx}
\usepackage{cleveref}
\usepackage{adjustbox}

\usepackage{xspace}

\newcommand{\numDataphys}{250\xspace}

\newcommand{\numTotal}{137\xspace}
\newcommand{\numPaper}{75\xspace}
\newcommand{\numShortPaper}{17\xspace}
\newcommand{\numThesis}{4\xspace}
\newcommand{\numWeb}{27\xspace}
\newcommand{\numVideo}{13\xspace}
\newcommand{\numAcademic}{96\xspace}
\newcommand{\numArtist}{37\xspace}
\newcommand{\numProfessional}{4\xspace}

\newcommand{\numActive}{47\xspace}
\newcommand{\numPassive}{79\xspace}
\newcommand{\numAugmented}{13\xspace}

\newcommand{\numAR}{3\xspace}

\newcommand{\numInfoVis}{27\xspace}
\newcommand{\numSciVis}{33\xspace}

\newcommand{\numSculpture}{85\xspace}
\newcommand{\numArtistic}{10\xspace}

\newcommand{\numSimplifying}{28\xspace}
\newcommand{\numSelf}{16\xspace}
\newcommand{\numLearning}{8\xspace}
\newcommand{\numResearch}{12\xspace}
\newcommand{\numAccessibility}{5\xspace}
\newcommand{\numSpatial}{8\xspace}
\newcommand{\numInstallation}{15\xspace}

\newcommand{\numPersonal}{13\xspace}
\newcommand{\numOtherData}{48\xspace}

\newcommand{\numPrint}{60\xspace}

\newcommand{\numSLS}{11\xspace}
\newcommand{\numCNC}{10\xspace}
\newcommand{\numLaser}{20\xspace}
\newcommand{\numHybridFab}{13\xspace}
\newcommand{\numAssembly}{74\xspace}

\usepackage[utf8]{inputenc}

\usepackage{graphics}
\usepackage{booktabs, multirow} 
\usepackage{soul}
\usepackage[table]{xcolor} 
\usepackage[utf8]{inputenc}
\usepackage{booktabs}

\usepackage{caption}
\usepackage{refstyle}
\usepackage{tabularx}
\usepackage{longtable}
\usepackage{xltabular}
\usepackage{bigstrut}
\usepackage{rotating}
\usepackage{lscape} 
\usepackage{color, colortbl}

\begin{document}


\maketitle
\begin{abstract}
    Physical representations of data offer physical and spatial ways of looking at, navigating, and interacting with data.  While digital fabrication has facilitated the creation of objects with data-driven geometry, rendering data as a physically fabricated object is still a daunting leap for many physicalization designers. Rendering in the scope of this research refers to the back-and-forth process from digital design to digital fabrication and its specific challenges. We developed a corpus of example data physicalizations from research literature and physicalization practice. This survey then unpacks the \say{rendering} phase of the extended InfoVis pipeline in greater detail through these examples, with the aim of identifying ways that researchers, artists, and industry practitioners \say{render} physicalizations using digital design and fabrication tools.  

-------------------------------------------------------------------------


\end{abstract}


\section{Introduction}
\label{sec:intro}

\begin{figure*}[t]
    \centering
    \includegraphics[bb=0 0 1263 460, width=\textwidth]{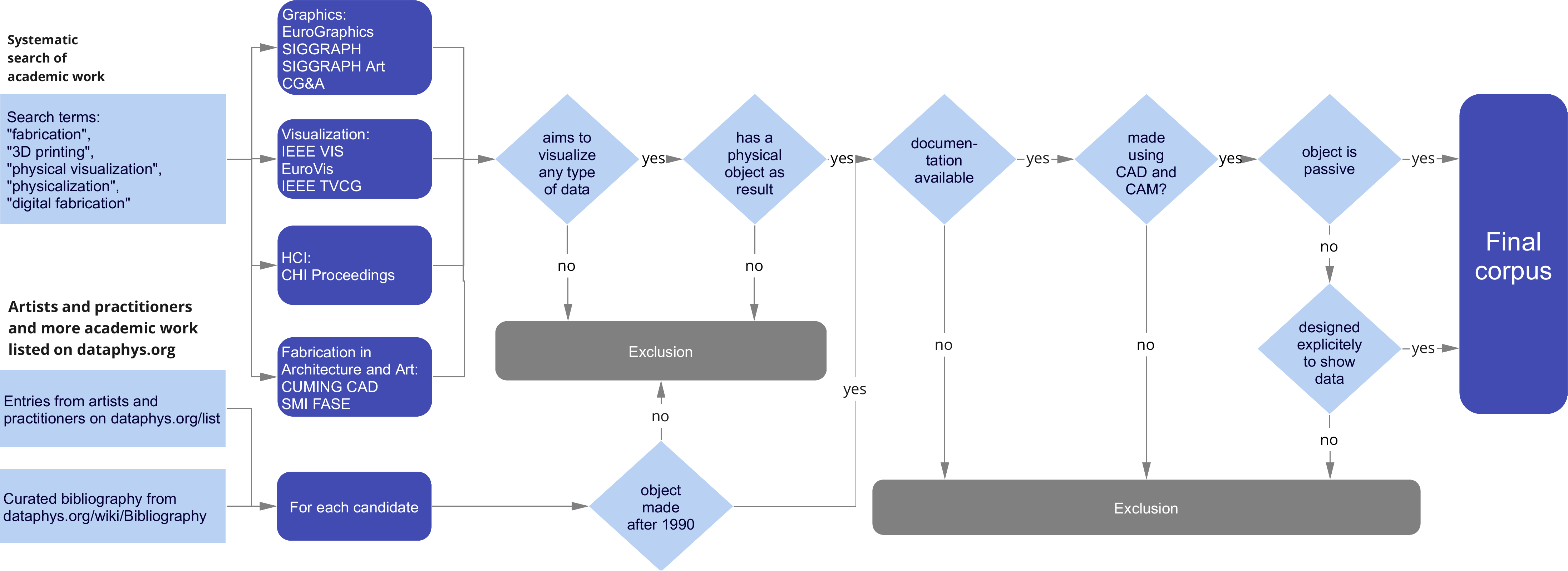}
    \caption{Decision graph for the curating process of our corpus.}
    \label{fig:corpus-curation}
\end{figure*}

Long before the invention of writing, people have used physical forms to record information\cite{Blombos}. 
Physical data representations --also called \textit{physicalizations}-- display data through the geometric or physical properties of an artifact~\cite{jansen2015opportunities}. 
Physicalizations are capable of leveraging perceptual exploration skills~\cite{jansen2015opportunities} to help users understand, explore, and perceive data. Research has shown that physicalizations can improve the efficiency of information retrieval and memorability of data when compared to similar designs shown on flat screens~\cite{jansen2013evaluating, stusak2015evaluating}; they can also positively impact data perception and exploration~\cite{taher2016investigating}, and they facilitate authoring of data representations for visualization novices~\cite{huron2014constructing, wun2016comparing}. Physicalizations inherit all of the practical and social advantages of everyday objects: they can be explored through touch, carried around, or possessed\cite{moere2008beyond}, and they can be directly manipulated\cite{taher2016investigating}. 
Data physicalization has both historic and contemporary applications in many domains, including geospatial visualization (e.g.,~\cite{djavaherpour2017physical, allahverdi2018landscaper}), planning (e.g.,~\cite{fetterman2014luminocity, aitcheson20053axis}), medicine (e.g.,~\cite{bucking2017medical, hadeed2018cardiac}), mathematics (e.g.,~\cite{segerman20123d, knill2013illustrating}), and education  (e.g.,~\cite{carroll20173d, higman2017hands}).

Designing and producing physicalizations requires expertise in both visualization and physical fabrication. Making precise physical objects that reflect data, such as architectural scale models, has historically been time-consuming, laborious, and costly. While today's advanced and accessible digital fabrication technologies have facilitated the process of physical fabrication from digital data, rendering data into a physicalization is still challenging. While digital fabrication machines have taken over the process of shaping some material into the desired form, diverse knowledge is required from preparing the design files for the machines to considering possible interactions between a chosen design, materials and fabrication techniques. We call the steps involved in this process the \emph{physical rendering process}. 

Physical rendering --or \textit{rendering}-- makes the visual presentation perceivable by bringing it into existence in the physical world \cite{jansen2013interaction}.
This transformation of data through rendering is not often a simple, straightforward process. 
Limitations of the fabrication's technology (e.g. size, speed and colour limitations) impose some restrictions in the transformation.
Physical rendering requires an interdisciplinary understanding of how data is represented and visualized (Visualization and Computer Graphics), how to design and create physical objects (Design and Fabrication), and how people physically interact with that data (Human-Computer Interaction).


In this survey, we focus on the rendering phase of the extended Infovis pipeline~\cite{jansen2013interaction} and review approaches and methodologies for converting data into digitally-fabricated physicalizations.
This STAR aims at addressing the following questions:
\begin{itemize}
    \item What is the target dataset and the resulting visualization \textit{idiom}, i.e., the distinct approach to create and manipulate the visual representation\cite{munzner2014visualization}?
    \item What are the dominant strategies/approaches towards physical rendering?
    \item What are the challenges of rendering transformation?
\end{itemize}

Our goal is to provide physicalization researchers and designers with a review of alternative physical rendering methods and their trade-offs, such that they can select rendering methods tailored to their goals and expertise. Although there exist other survey papers related to various fabrication approaches\cite{Hullin2013Computational,bermano2017fabrication-aware,livesu20173d}, their focus is not on physical rendering which requires a systematic exploration of rendering methods.

 In this report, we detail our methodology (Section \ref{sec:methods}) for gathering our sample of physicalization papers and examples. We then describe our classification approach (Section \ref{sec:classification}), which includes  InfoVis vs. SciVis, pragmatic vs. artistic, and passive vs. active. Section \ref{sec:classification} also covers two more approaches toward the classification of physicalizations: one from an application point of view and one from an idiom point of view. 

We then discuss our findings from our analysis. In Section \ref{sec:data}, we discuss the range of datasets used as target data to make physicalizations by different communities and practitioners. We review digital design tools and methods, digital fabrication tools and technologies, and approaches to building augmented and active physicalizations in Section \ref{sec:rendering}. 
Finally, we describe the rendering process in greater detail
(Section \ref{sec:discussion}). 
These challenges include decisions made during design and fabrication that have implications for how data is represented. 
This section also discusses the role of iterative design and usability testing as part of how we refine the design of physicalizations. 
Ultimately, these challenges reflect opportunities and directions for future research.



\section{Methodology}
\label{sec:methods}

In this section, we discuss how we assembled our corpus of physicalization examples for analysis. 

\subsection{Assembling Corpus of Physicalization Examples}

Many academic and art communities explore the physicalization of data. We built a corpus from two sources: (a) a systematic literature search and (b) specific physicalization examples from dataphys.org

Our systematic literature search started by filtering papers, short papers, and posters published between 2010 and 2020 that met a keyword search (CAD, modelling, data design, data-enabled design, data-driven design, CAM, fabrication, 3D printing, computational manufacturing, digital fabrication, physical visualization, physicalization, data materialization, embodied interaction, installation, physical, physical material, prototype, rapid prototyping, shape-changing, spatialization, tactile, tangible, tangible user interfaces, wearable, actuation, personal data) in the following academic communities: 
\begin{itemize}
\item Computer Graphics (Eurographics, SIGGRAPH, SIGGRAPH Asia, IEEE CG\&A) 
\item Visualization (EuroVis, IEEE Vis, IEEE TVCG), 
\item Human-Computer Interaction (CHI Proceedings) 
\item Fabrication in Art and Architecture (SIGGRAPH Art, SMI FASE, CUMINCAD). 
\end{itemize}



\begin{figure*}[t]
    \includegraphics[bb=0 0 1263 335, width=\textwidth]{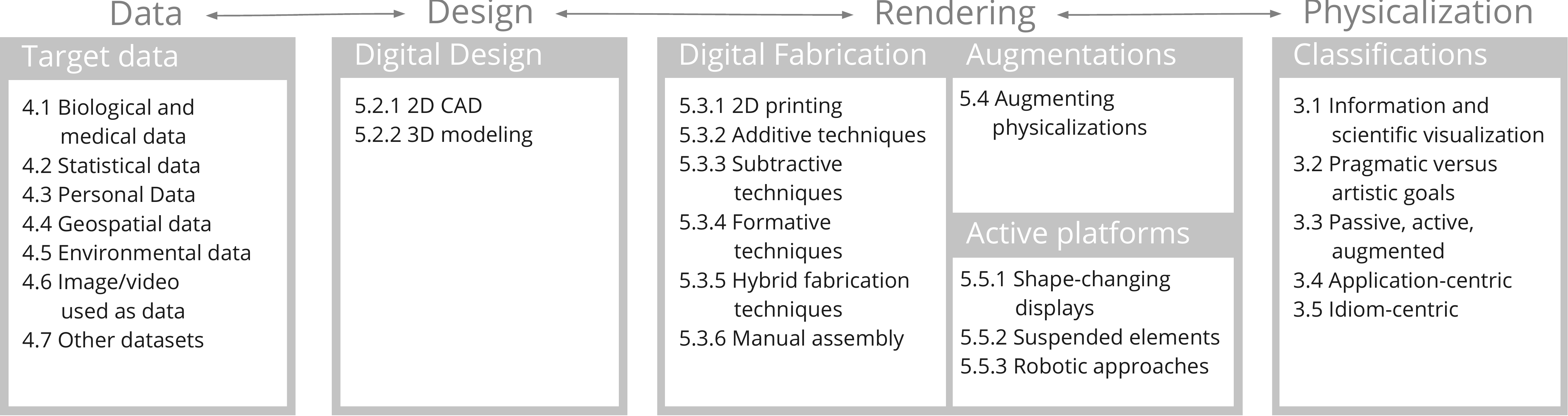}
    \caption{Physical Rendering Pipeline with digital fabrication, presenting the main sections of the paper and techniques to digitally fabricate a physicalization. Each square represents a main step, sections can be identified below their name in Italic.}
    \label{fig:data-phys-pipeline}
\end{figure*}

Meanwhile, we wanted to also include examples from the broader art and design community whose physicalizations may not appear in academic literature. Dataphys.org has actively collected examples of physicalizations from various disciplines since 2013. 
We excluded work from before 1990 as CAD/CAM technologies were less common. Also, we only considered examples with proper documentation (e.g. published papers and reports).  


In the end, we gathered \numDataphys examples, for which we filtered out the entries that did not have any available documents explaining their physicalization process. This study was also quite helpful in making us more familiar with different communities working on physicalizations.

Once we established this initial corpus of data physicalization examples from academic and practitioner communities, we continued to filter based on (a) availability of quality documentation with adequate detail to address research questions and (b) the use of digital design (CAD) or fabrication (CAM) software and tools. We then looked at whether the physicalization was a passive object, or represented through an active physical platform.  We excluded any active physical platforms that did not have specific data physicalization applications designed for them.


 A summary of the paper collection and corpus curating process is presented in \autoref{fig:corpus-curation}. Our final sample includes \numTotal works – \numPaper long papers, \numShortPaper short papers and posters, \numThesis thesis and dissertations, \numWeb works presented on websites,  \numVideo videos. 
 \numAcademic physicalizations are designed and developed by academic groups and researchers, \numArtist projects are made by artists and practitioners, and the professional community, such as architects, were also part of the physicalization community by making \numProfessional projects. Our corpus and its analysis are available to readers as static tables included in the paper (see \autoref{tab:Taxonomy1} and \autoref{tab:Taxonomy2}), as well as an interactive online version under \url{https://yvonnejansen.github.io/physicalization-rendering/}.

\subsection{Analysis}

The process of physicalization includes some actions and activities such as collecting data from different types, digitizing data and convert it to a visual form, fabrication, etc. We structure the main categories of our coding schema into a process pipeline of physical rendering during the fabrication of a physicalization in \autoref{fig:data-phys-pipeline}. This pipeline represents some of the possible steps coded in the collection to digitally fabricate a physicalization, from the data to the final artefact. Each step represents a section of the paper.


\section{Physicalization Classification Schemes}
\label{sec:classification}

Our corpus contains a breadth of physicalizations that have emerged from different communities (research, art, design) with diverse skill sets, intentions, and approaches to physicalization. 
In this section, we discuss the breadth of our corpus along with several factors: information and scientific visualization; pragmatic and artistic; passive, active, and augmented physical objects.  We also discuss application-centric and idiom-centric classifications of these physicalizations. Note though that categories within these factors and classification schema are not mutually-exclusive, and some physicalizations can be described as simultaneously addressing multiple categories.


\subsection{Information and Scientific Visualization}
\label{subsec:infovis}


Physicalizations can be categorized by a classic method of classifying visualizations: distinguishing between Information Visualization (InfoVis) and Scientific Visualization (SciVis). This distinction is, however, elusive, difficult to define, and controversial within the visualization community. One definition of the distinction between InfoVis and SciVis by Tamara Munzer: \say{it's InfoVis when the spatial representation is chosen, and it's SciVis when the spatial
representation is given}~\cite{munzner2008process}. 

Following this definition, our corpus includes \numInfoVis papers and projects that can clearly be categorized as Infovis and \numSciVis that can clearly be categorized as SciVis (see \autoref{tab:Taxonomy1} and \autoref{tab:Taxonomy2}).  
Both categories tend to not focus on specific types of data and include a wide variety of examples.  InfoVis physicalizations in our corpus include the representation of country indicators to explore correlations between data series~\cite{dwyer2004two}, personal activity data~\cite{stusak2014activity}, or time-series data of one's baby as a personal memento (e.g.,~\cite{swaminathan2014supporting}). Examples of SciVis physicalizations include physical maps to form connections between themes in a gallery space\cite{priestnall2012projection} or to explain the future of energy in Hawaii~\cite{kirshenbaum2020data}), to explore volumetric, anatomical data~\cite{nadeau2000visualizing} or 4D MRI blood flow data~\cite{ang2019physicalizing}), to understand the assembly of macro-molecules or viruses using passive physical models~\cite{bailey1998use} or combined with augmented reality overlays~\cite{gillet2005tangible}.



\subsection{Pragmatic vs. Artistic Goals}
\label{subsec:pragmatic}
We also looked at whether a physicalization was created in pursuit of pragmatic or artistic goals.
We adopted Robert Kosara's interpretation of pragmatic visualizations as having ``the goal [...] to explore, analyze, or present information in a way that allows the user to thoroughly understand the data'' and of artistic visualizations as having the goal ``to communicate a concern, rather than to show data''~\cite{kosara2007visualization}. 
In our classification, we considered physicalization examples representing data in a playful manner, to express concerns, or to offer inspiration as artistic, regardless of whether or not they were made by artists.

However, the distinction between pragmatic and artistic physicalizations is blurry. Examples like \emph{a piece from the pie chart}~\cite{rust2014piece}, a robotic pie-charts-on-pies machine, uses a classical encoding (pie charts) in an art exhibition with the intent to draw attention to gender distributions in the tech world. This example is simultaneously pragmatic (allowing the viewer to thoroughly understand the data) and artistic (made with the intent to communicate a concern). Our corpus includes around \numArtistic artistic physicalizations and \numSculpture data sculptures listed in the \url{dataphys.org/list} site as well as the SIGGRAPH Art track. Many of these examples are pragmatic as well.

 \subsection{Passive, Augmented, and Active Physicalizations}
 \label{subsec:passive}

Another dimension on which physicalizations can be classified is how they employ computational components. Many physicalizations are disconnected from all types of computational machines once fabricated. We call these physicalizations \emph{passive} in line with previous work~\cite{jansen2014physical}. Note that \emph{passive} only refers to the use of computational power and not to the support of interactivity more generally. We discuss in Section \ref{subsec:assembly-interaction} how different fabrication and assembly techniques can permit different levels of (manual) interactions such as sorting and filtering~\cite{jansen2013interaction}. Our corpus includes \numPassive examples of passive physicalizations.

In \numAugmented examples, we observed the combination of passive physicalizations with augmentations such as projections or augmented reality overlays which provide access to computational functionality on some of the data dimensions. For example, Gillet and colleagues~\cite{gillet2005tangible} presented physical molecule models where users can explore the interaction of their electromagnetic fields in augmented reality when the molecules are brought close together. In another example, Hemment and colleagues~\cite{hemment2013emoto} augmented physical height maps of Twitter sentiments about the 2012 Olympic Games by projecting on top of them and thus enabling visitors to highlight different aspects of the data interactively. We discuss augmented physicalizations in more detail in Section \ref{subsec:augment}.

Finally, we identified \numActive examples of physicalizations that are dependent on some form of computational or at least electrical power to show their data to an observer. There are many different ways of realizing this which we review in Section \ref{subsec:active} in more detail. Using active rendering techniques not only enables the addition of some computer-supported interactions -- as with augmented physicalizations -- but also supports functionalities such as updating or loading different data sets (e.g., \cite{houben2016physikit, taher2015exploring, LeGoc2018dynamic, follmer2013inform}. However, active physicalizations tend to suffer from scalability issues: generally, one actuator is required per data point and adding more actuators to an already existing system to accommodate a larger data set can prove difficult.

\subsection{Application-centric Classification}

One possible method of classifying physicalizations is through different applications that they can be used for. For instance, some physicaliztaions are designed to \textbf{simplify the understanding of information or scientific data} and help a specific group of practitioners or general public easier understand such concepts. Such physicalizations raise awareness, help in making better decisions, and can be used as collaboration tools among various professional or academic groups (e.g.,\cite{thrun2016visualization,priestnall2012projection,ang2019physicalizing,kirshenbaum2020data,LeGoc2018dynamic}). \numSimplifying works on our corpus belong to this category of physicalizations.

Another group of physicalizations in an application-centric classification are the works that are made to aid people in keeping track of various tasks and activities in their personal lives and \textbf{raise self awareness} (\numSelf works in our corpus). Many of such physicalizations focus on personal activity and health tracking data that we will discuss in Section \ref{subsec:PersonalData}. Another goal for making such physicalizations has been keeping track of progress during PhD studies\cite{Schneider2012Lego,swaminathan2014supporting}.


Physicalizations have a great potential for \textbf{improving accessibility}, such as tools for helping people with limited or no vision (e.g.,\cite{patel20173d,tymms2018quantitative,swaminathan2016linespace}) (\numAccessibility total examples). They can also be used for \textbf{learning and education} (e.g.,\cite{djavaherpour2017physical,bader2018making} (with \numLearning total examples), as \textbf{research and engineering} tools (e.g.,\cite{miyashita2016zoematrope,vsimbelis2014metaphone} (with \numResearch total examples), and for \textbf{presurgical planning} (e.g.,\cite{bader2018making}).

\subsection{Representational Idioms of Physicalizations}
\label{subsec:idiom}

Munzner calls every distinct approach to create and manipulate a visual representation from the abstract data an \textit{idiom}\cite{munzner2014visualization}. She introduces two major categories in idiom design: visual encoding idiom, i.e., representational idiom, and interaction idiom. The visual encoding idiom controls what people see in a visualization. 


Based on the physicalizations reviewed in our corpus, a high-level categorization of representational idioms can be introduced as follows: physical charts, topography and elevation models, informative spaces and installations, and unique data objects. 


\noindent\textbf{Physical Charts.} Munzner's visual encoding idioms reflect different graphical chart types (e.g., bar charts, line graphs, etc.). Many physicalizations extend visual encoding idioms from graphical representation into physical 3D objects. These include physical bar charts (e.g.,\cite{swaminathan2014supporting,jansen2013evaluating}), pie charts (e.g.,\cite{fens2014personal}), scatterplots (e.g., scatterplots on an extruded 3D map of NYC\cite{Kauffman2013NYC}), and prism maps (e.g.,\cite{kane2014tracking,swaminathan2014supporting}). 

\begin{figure}
  \centering
   \includegraphics[bb=0 0 1161 548, width=\linewidth]{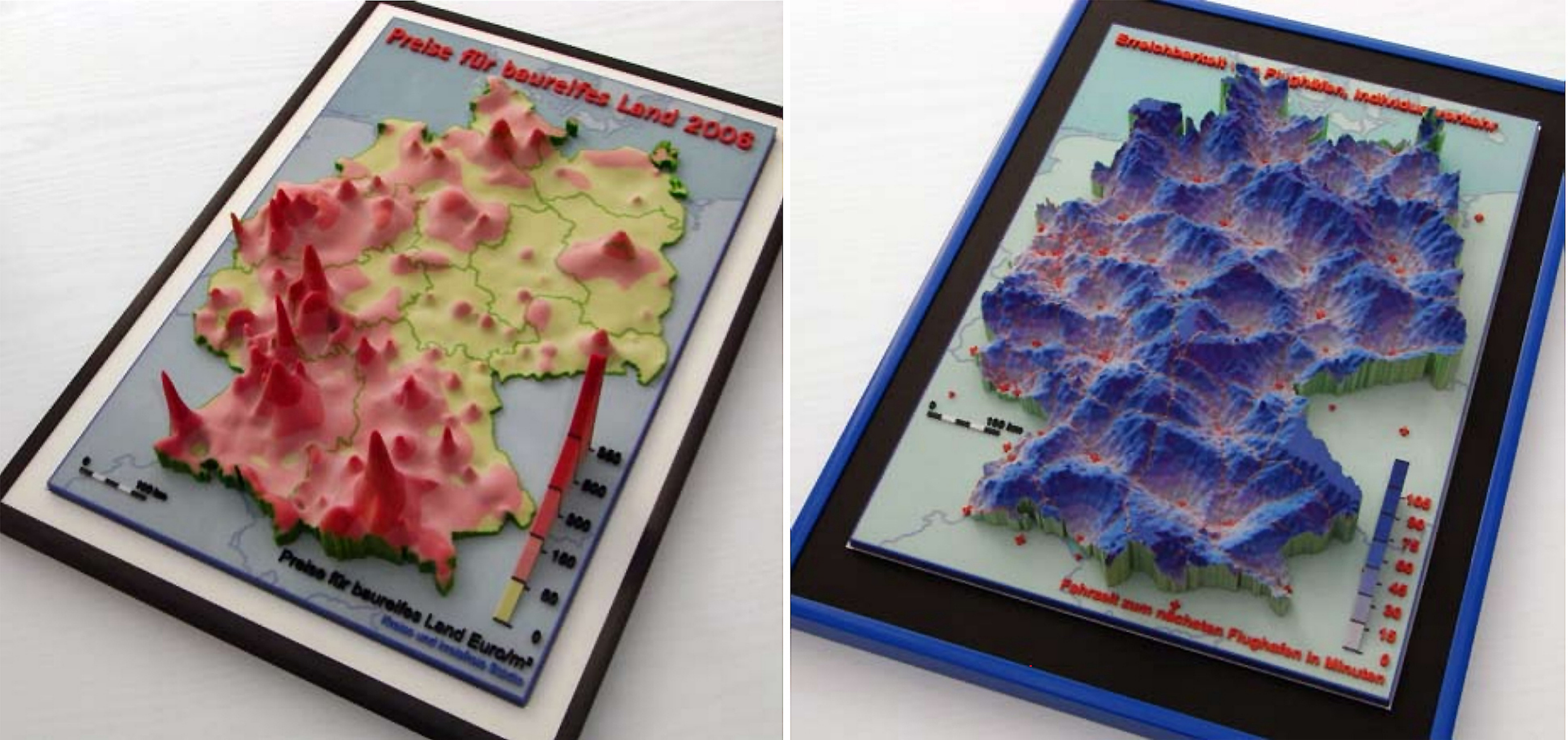}
   \caption{Using Physical Elevation Models for showing the average prices for building lots in Germany (Left) and time-distance to the next airport (Right). Images taken from~\cite{rase2011creating}.}
           \label{fig:PEM}
\end{figure}

\noindent\textbf{Topography and Physical Elevation Models.} Physical Elevation Models generally physicalize elevation data, terrains and topographies (e.g.,\cite{tateosian2010tangeoms,nittala2015planwell}).  They include relief models (e.g.,\cite{priestnall2012projection}) or terrain models that are used as a base for other physicalizations, such as airplane trajectories\cite{Solid2003Airplane}. However, in many cases, the physical characteristics of topography surfaces, including height the heightmap, show datasets other than elevation and topography. In such cases, the surface heights of the elevation model are proportional to data,  resulting in a smooth interpolated surface (refer to Section \ref{subsubsec:3DCAD} for more details about modelling surfaces). For instance, Rase made physical elevation models to show average prices of building lots in Germany\cite{rase2011creating} (see \autoref{fig:PEM}); Gwilt et al. mapped package openability data to surface roughness\cite{gwilt2012enhancing}. 


\noindent\textbf{Informative Spaces and Installations.} These physicalizations are mostly architectural spaces or artistic installations, designed with data, for the purpose of conveying a message. Architects and designers now use computational design methods to leverage available data streams and generate novel forms and spatial opportunities\cite{brown2017designing,gonzalez2017space}. Physicalizations with this representational idiom aim to provide an atmospheric experience for users while reflecting a message from their target data. In such work, communicating information and producing abstract effects (e.g., with lights, colours, movements) are mixed in the form of an installation at an architectural scale (i.e., to form a space). Such approach helps in mixing the didactic and literal representations with qualitative and atmospheric experiences. Didactic spaces are also referred to as \textit{data spatialization}\cite{marcus2014centennial}. For instance, \textit{Data-spatialized Pavilion}\cite{hosseini2019data} introduces a novel method to make a data-driven pavilion through catoptric (mirror-assisted) anamorphosis, where the input data defines the physicality of the pavilion and simultaneously remains readable. In another example, \textit{Weather Report}\cite{keefe2018weather} uses a set of two illuminated balloon walls, one for representing real-time weather data (quantitative) and one for visualizing the audience's memories of weather (qualitative). 
There are \numSpatial examples of informative spaces in our corpus and \numInstallation examples in the form of active installations.  



\noindent\textbf{Unique Data Objects.} Unique objects designed with data -- frequently referred to as data sculptures-- can take many forms, shapes, and scales. Many of the physicalizations in our corpus are objects small enough to be picked up and held. For example, \emph{Motus Forma} shows 10 hours of movement trajectories in the lobby area of Pier 9 \cite{Allen2016Motus}; Doug McCune's physical maps show data relating to living conditions in San Francisco~\cite{McCune2013Maps,Mccune2016Housing}; Loren Madsen's data sculptures represent the increase of cost of living from 1960 to 1994~\cite{Madsen1995Early}. Some physicalizations were created as wearable clothing~\cite{Perovich2014Data, Chen2014Xpose} or jewelry~\cite{Kang2017Wearable,lee2015patina}. Some artists have taken unique approaches to make data physical. For instance, the \textit{Snow Water Equivalent Cabinet} shows snowpack measurements of the years 1980-2010 by making a drawer-like plywood sculpture, where the size of each drawer corresponds to the annual precipitation by year~\cite{Adrien2011data}.

\begin{figure}
  \centering
   \includegraphics[bb=0 0 1610 458, width=\linewidth]{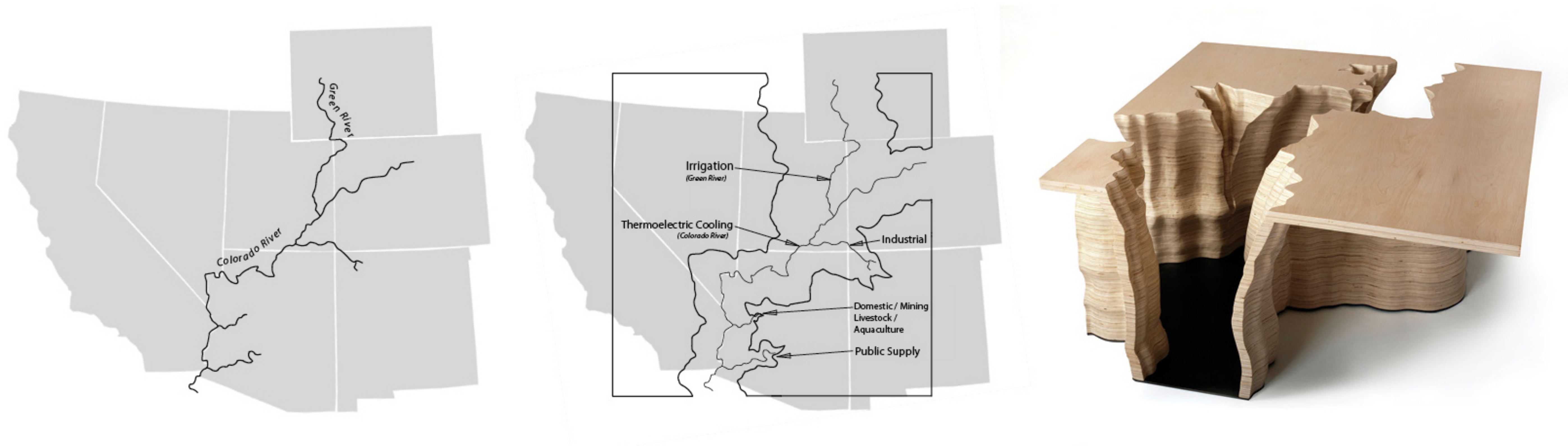}
   \caption{Physicalization designers sometimes look to nature for inspiration. This data sculpture by Adrien Segal, shows trends in water use and uses a map of the Colorado river as a design concept. Images taken from the sources in\cite{Adrien2011data}.}
           \label{fig:trends}
\end{figure}



\section{Target Data for Physicalization}
\label{sec:data}




Many types of datasets have been transformed into physicalizations, from personal activity data~\cite{khot2014understanding,stusak2014activity,lee2015patina} to thesis progress data~\cite{schneider2015visualisation}. In this section, we provide an overview of various types of datasets and data types that have been represented in physicalizations.

We note that there is overlap between these categories --  datasets from other categories may be represented in statistical forms, personal data can be geospatial or about personal health or medicine.


\begin{figure*}[t]
    \centering
    \includegraphics[bb=0 0 3249 573, width=\textwidth]{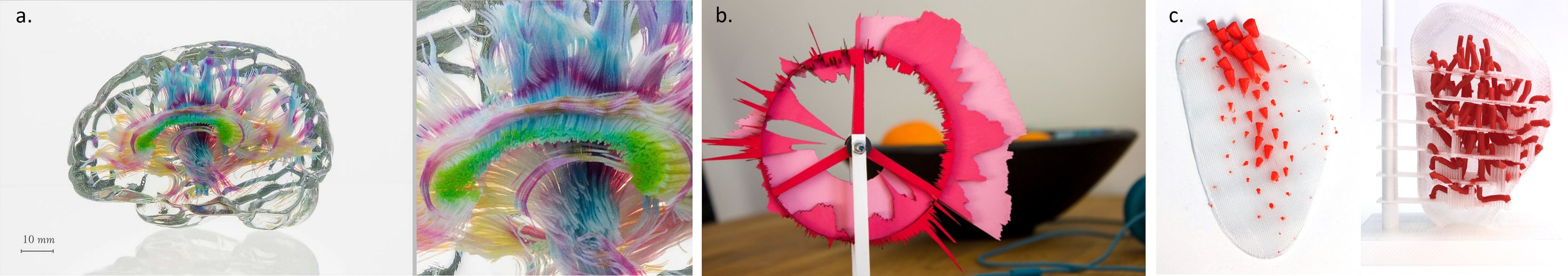}
    \caption{Examples of biological and medical data physicalizations. (a) Diffusion-weighted MRI data capturing the diffusion of water molecules in white matter brain tissue. Image taken from\cite{bader2018making}, (b) Wooden display showing heart rate and skin temperature. Image taken from\cite{fens2014personal}, (c) Cardiac blood flow data shown with slices of glyphs and cones. Images taken from\cite{ang2019physicalizing}.}

    \label{fig:biology}
\end{figure*}

\subsection{Biological and Medical Data}
\label{subsec:biological}

Due to the complexity and delicacy of medical and biological datasets, tangible visualizations that can show different modes of such datasets for a range of stakeholders can be quite useful. 
As a result, physicalization for these datasets has been broadly studied and practiced.

In \cite{gillet2004computer,gillet2005tangible}, Gillett et al. combine 3D printing and virtual reality to improve learning complex biological molecule structures; using their system, people manipulate a physical 3D printed model that is tracked by a camera, controlling the viewpoint of a graphical visualization displayed on a screen. 
Rezaeian and Donovan represented the personal DNA data of individuals as 3D printed jewelry\cite{rezaeian2014design}. 
Variety of datasets including MRI has been 3D printed in plausible forms using multimaterial voxel-printing method in various colors (see \autoref{fig:biology}a) \cite{bader2018making}.
Metaphone \cite{vsimbelis2014metaphone} turns individual's bio-data (e.g., Galvanic Skin Response (GSR) and Heart Rate (HR)) into a colorful 2D painting.
Personal health data is physicalized in \cite{fens2014personal} through a multimodal representation. For instance, a two dimensional
wooden radial display that simultaneously visualizes temporal heart rates and skin temperature (see \autoref{fig:biology}b). 
Nadeau and Bailey created 3D physical models with interlocking pieces from medical volumetric data via solid free-form fabrication equipment \cite{nadeau2000visualizing}. 
Thrun and Lerch used 3D printing to represent high-dimensional datasets such as pain phenotypes as a landscape in four different colors (i.e., white, red, green, blue, yellow), highlighting distance \cite{thrun2016visualization}. 
Ang et al.~\cite{ang2019physicalizing} physicalized blood-flow datasets by 3D printing slices of curves or glyph to resemble flow directions in a volume (see \autoref{fig:biology}c).
Lozano-hemmer physicalized viewers' heart rates with a set of light bulbs hanging in a room, synchronizing the bulbs with each heart rate as viewers began interacting with the work~\cite{Lozano2006PulseRoom}. 

Geurts and Guglielmetti \cite{geurts2015imagining,geurts2018imagining} discussed the possibility of capturing thoughts and the relationship of cognitive and emotional to one's work and living environments in digital and visual forms (e.g., images). Neural connections in the brain are simulated and physicalized by a set of bottles spinning on a table forming various patterns~\cite{Lozano2004Synaptic}.
To promote physical activities, EdiPulse~\cite{khot2017edipulse} transformed self-monitored physical activity data into chocolate treats that get 3D printed to produce a specific icon or message.

Biological data has also been physicalized via sonification -- the production of sounds based off of data. Barrass used a head-related transfer function (HRTF) to generate a bell-shaped 3D physical model \cite{barrass2011phsyical,barrass2012digital} and to transform blood pressure data into a singing bowl \cite{barrass2014acoustic}.

\subsection{Statistical Data}
\label{subsec:math}

Engaging physicalizations can be very helpful for communicating statistical datasets with the audience \cite{jansen2013evaluating,marcus2014centennial}.
Statistical datasets are usually quantitative values represented in numerical or string formats. Examples of such datasets include water consumption (in million gallons per day)~\cite{Adrien2011data}, class sizes and the number of graduates~\cite{marcus2014centennial}, etc.  Here, we list specific examples of statistical and mathematical datasets from our sample.

\begin{figure*}[t]
    \centering
    \includegraphics[bb=0 0 5066 708, width=\textwidth]{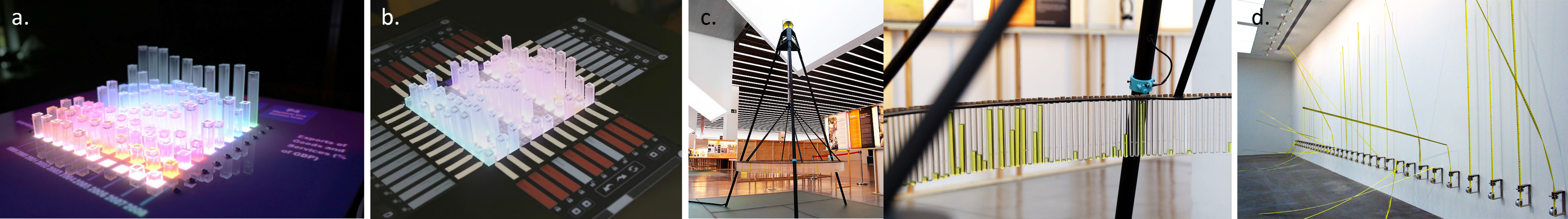}
    \caption{Engaging physicalizations help significantly in conveying the message of mathematical and statistical datasets: (a) Actuating physical bar chart with LEDs to show international export data. Image taken from\cite{taher2015exploring}, (b) A dynamic shape-changing display showing ratings from inhabitants of 46 European countries on topics such as social issues, politics, military, healthcare, and economy. Image taken from\cite{taher2016investigating}, (c) Drip-By-Tweet shows real-time physicalization of votes collected through Twitter. Images taken from\cite{Domestic2014Drip}, (d) Motorized measuring tapes physicalizing the time each viewer spends in the installation. Image taken from\cite{Lozano2011Tape}.}

    \label{fig:statistical}
\end{figure*}

Taher et al. created responsive bar charts to communicate statistical data (e.g., international export data) with  rods and RGB LEDs \cite{taher2015exploring,taher2016investigating} (see \autoref{fig:statistical}a,b). 
Pulse~\cite{Ferrara2012Pulse} is a tangible line graph composed of a string whose position is modified by six servo motors. 
Drip-By-Tweet \cite{Domestic2014Drip} visualizes the statics related to a voting mechanism collected on Twitter by a series of tubes whose amount of fluid changes based on the number of cast votes (see \autoref{fig:statistical}c). 
In Tape Recorders~\cite{Lozano2011Tape}, motorised measuring tapes visualize the amount of time that visitors spend in a particular installation (see \autoref{fig:statistical}d).
Kauffman and Brenner~\cite{Kauffman2013NYC} created a physicalization of high school drop outs in New York by highlighting the locations of schools on the map with a set of beads. The beads are connected to a string below with lengths relative to the number of students who dropped out.

To raise awareness about the lack of female representation in art and tech, \emph{A Piece of the Pie Chart} transformed gender ratios into real, edible pie charts \cite{rust2014piece}. 
Floating charts~\cite{omirou2016floating} is an acoustic levitation display for placing free-floating objects that has been constructed to visualize a dynamic floating chart to reflect changes in data. 

Le Goc et al.~\cite{LeGoc2018dynamic} introduced Zooids, a dynamic physicalization where small moving robots form patterns and clusters representing data points to facilitate decision making (e.g., ranking applicants for departmental admissions). Emoto~\cite{hemment2013emoto} used origami-like data sculptures to communicate Twitter data related to London 2012 Olympics events. Fantibles~\cite{khot2016fantibles} is a personalized memorabilia capturing an individual's commentary about sports (e.g., cricket) through a nested double-ring physicalization.

Starrett et al.~\cite{starrett2018data} turned the famous computer graphics object, Utah teapot, into a visualization by changing its base to a curve representing datasets by intersecting circles.  Chaotic Flow~\cite{Lukassen2012Chaotic} is an installation of colorful flowing liquid that visualizes the flow of Copenhagen bikes. Perovich et al. fabricated lace patterns for clothes based on air pollution datasets~\cite{Perovich2014Data}. 
McCune created physical maps physical thematic maps to turn \say{horrible data} (e.g., murders or natural disasters) into visually pleasing physicalizations~\cite{McCune2013Maps}. 
Cosmos~\cite{Jarman2014Cosmos} is a spherical wooden sculpture that represents data from forests that describe the take-up and loss of carbon dioxide by trees. 
Data Moir\'{e} \cite{Huang2017moire} is an effort to physicalize the data on IBM Digital Analytics Benchmark to a large-scale feature wall that is CNC-machined. Madsen also represented the evolution in the world population from 10,000 BCE to today as a 20-meter long data sculpture~\cite{Madsen1995Early}.


Radically different materials and forms have been used for math dataset physicalizations such as crystal engraving \cite{bourke2015novel} or paper \cite{DeMarco2011Paper}. 
For instance, to facilitate students with visual impairment to learn math, VizTouch has been developed to produce 3D printed tactile visualizations to represent mathematical contents such as graphs \cite{brown2012viztouch}. 
Wavefunction \cite{Lozano2007Wave} uses a set of chairs (50-100) that are arranged like a regular array of rows. The height of these chairs change when an audience approaches a chair producing a crest and the height change propagates through other chairs.



\subsection{Personal Data}
\label{subsec:PersonalData}

Self-monitoring practices raise awareness about an individual's personal habits; as a creative representational method, physicalizations can encourage different groups of people to actively monitor their progress and become conscious about their habits and behaviors, such as physical activity\cite{khot2014understanding}. Towards this goal, Stusak et al. designed a system that collects datasets from users' running activity (e.g., duration, distance, elevation gain, average speed) and generates multiple types of activity sculptures~\cite{stusak2014activity}. The 3D printed sculptures-- a jar, a necklace, a lamp, and a figure-- were delivered to users as personal tokens (see \autoref{fig:Personal}b). \textit{Patina Engraver} uses the gradual development of patinas to map user activity data to a wearable wrist band by applying stippling technique (i.e., a technique that creates a pattern simulating varying degrees of solidity or shading using small dots)\cite{lee2015patina}. Personal activity and sleep data have also been used to make personalized jewelry and fashion items (e.g.,\cite{Kang2017Wearable,hakkila2016aesthetic}). In an interesting data-driven design approach, Nachtigall et al.\cite{nachtigall2019encoding} personalized the design of a pair of shoes by encoding the footsteps data of their owner.    

Some personal physicalizations were designed to keep people motivated. For instance, \textit{TastyBeats} prepared drinks for users after a workout, based on their heartrate values~\cite{khot2015tastybeats}. While people with heartrates in the low activity zone only received water, those who elevated their heartrate to the intense level zone were given a rich-flavored drink. In another approach to motivating activity through food, Khot et al. translated physical activity data into 3D printed chocolate treats~\cite{khot2017edipulse}. \textit{Go and Grow} motivated tracking and self-reflecting on their fitness data by mapping activity data proportionally to the amount of water given to a living plant; the more active the plant owner, the healthier their plants become~\cite{botros2016go}. 

With \numPersonal works in this category, physicalizations that reflect personal data show an emerging and interesting direction for further exploration. 
Moreover, the studies on personal physicalizations demonstrate how engaging idioms (food, plant growth, wearable objects) can encourage and motivate physical activity and provide pleasurable interactions with personal data. 
As a deeper investigation of the intersection of personal data and materiality, Khot et al.~\cite{khot2020shelfie} reviewed examples of personal physicalizations to propose a conceptual design framework for creating material representations of physical activity data.

\subsection{Geospatial Data}
\label{subsec:geospatial}

Geospatial datasets are well suited for fabrication as they refer to a particular spatial location or geographical scene. Therefore, many works benefitted from different physicalization approaches to better represent such datasets.

Geospatial datasets are typically of four main formats: imagery datasets (e.g., satellite images), elevation datasets (e.g., DEM), vector datasets (e.g., roads, boundaries), or 3D geometries (e.g., 3D buildings)~\cite{mahdavi2015survey}. Various forms of geospatial physicalizations have been developed for the purposes of education~\cite{kane2014tracking,moorman2020geospatial}, providing scenery models or data~\cite{rase2011creating}, or raising awareness~\cite{Kildall2014Water}.
In the following, we discuss such approaches and provide details about their methodology.

\begin{figure}
  \centering
   \includegraphics[bb=0 0 1256 445, width=\linewidth]{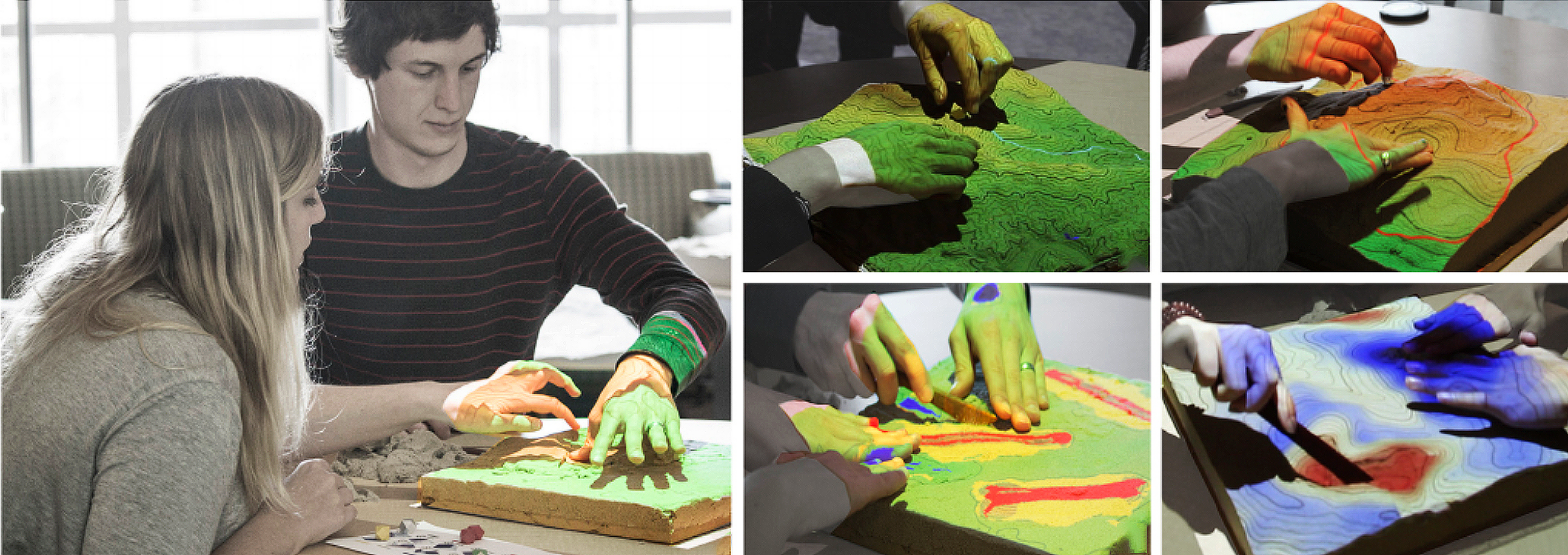}
   \caption{Tangible Landscape is a malleable model, equipped with projectors, that enables users learn about various aspects of topographical properties. Image taken from \cite{millar2018tangible}.}
   
   \label{fig:geospatial}
\end{figure}

Tangible Landscape is a 3D educational physicalization to teach topography (i.e., the shape of terrains)~\cite{millar2018tangible}.
Fabricated using molds, this physical landscape was a soft malleable model equipped with top-mounted projectors to provide feedback and guidance to users (see \autoref{fig:geospatial}). 
Benefiting from affordability and accessibility of 3D fabrication, PARM \cite{priestnall2012projection} was a tangible geographic display in which a projector reflects data on a CNC-machined base topography.
Landscaper~\cite{allahverdi2018landscaper} used interlocking colored pieces to physicalize a sizable landscape of a given region with various datasets.
Djavaherpour et al.~\cite{djavaherpour2017physical} 3D printed equal-area physical tiles for a globe, which served as placeholders for the corresponding regions on the Earth where additional datasets could be attached or layered.
Along the same line, Dadkhahfard et al.~\cite{dadkhahfard2018area} fabricated a curved equal area representation of the Earth on which various dynamic datasets were projected.
TanGeoMS~\cite{tateosian2010tangeoms} integrates a laser scanner, projector, and a flexible physical 3D model; end-users can control a digitally projected simulation by add and remove artifacts on the 3D model.
Created for military purposes, Xenotran~\cite{Schmitz2004Xenovision} is a self-reconfigurable solid terrain model whose surface movements are controlled by 7000 actuators. 

Geospatial physicalizations have also been used to address interesting applications: depicting a case study of a plane crash~\cite{Solid2003Airplane},  showing parks and forests in Berlin~\cite{Meier2017Green}, visualizing world population density~\cite{Badger2013Population}, and showing people movements in a lobby space~\cite{Allen2016Motus}.
In addition, we found examples of artistic geospatial physicalizations, such as the data-spatialized pavillion~\cite{hosseini2019data}, where a terrain model from top view displays a particular artistic feature (e.g., Mona Lisa painting).

\subsection{Environmental Data}
\label{subsec:environmental}

Environmental data addresses measurements of the environment, its systems, and impacts on its ecosystem. Engaging visualizations of environmental datasets is crucial to raise awareness about critical issues including wildfire, global warming, animal extinctions, etc. Many of these examples are produced with artistic goals to offer a critical perspective.
Segal transformed the amount of water stored as snow throughout a season into furniture, where the choice of forms and materials connected back to the origins of the data~\cite{Adrien2011data}.
Aweida~\cite{Aweida2013RobotArranges} combined robotics and art to build a physicalization of wind simulations via a foam board, a robot and a series of nails. Elsinki wind's travels is physicalized on a piece of wood by a CNC machine in Windcuts~\cite{Knapek2012Windcuts}. Whitelaw turned environmental datasets into artistic artifacts by making a bracelet from one year of weather data of Canberra and a measuring cup from monthly average temperatures in Sydney over 150 years~\cite{Whitelaw2009Weather}. 

Some environmental physicalizations, however, reflect data from an end-user's immediate environment. Physikit~\cite{houben2016physikit} was a series of physical ambient visualizations that let end-users to explore and engage with environmental data. Data from Physikit was] visualized through movement (PhysiMove), vibrations (PhysiBuzz), air (PhysiAir), and light cubes(PhysiLight).  Physicalization designers have also leveraged the biological properties of plants that respond to environmental conditions to create human-readable structures~\cite{yao2015biologic,velikov2014nervous}. Weather Report~\cite{swackhamer2017weather,keefe2018weather} visualized weather data by building a steel tube frame supporting an array of white balloons, on which weather data is projected as a color-coded animation.


\subsection{Image/Video Data}
\label{subsec:image}

We also found examples of people transforming images and video into physical artifacts in order to add tangibility or artistic features.
Zhao et al.~\cite{zhao2016printed} produce artistic lampshades projecting grayscale images onto surrounding walls.
String Art replicates an image by several straight lines of strings that are tied to a set of pins located on a frame~\cite{birsak2018string}. 
Portal~\cite{hosseini2020portal} is a structure produced by a laser cutter and a set of mirrors to create an image that does not exist in the environment by reflecting colors from another given image.
In addition, to produce paintings, watercolor woodblocks are designed to ease the process of producing several copies of a painting~\cite{panotopoulou2018watercolor}.
MoSculp~\cite{zhang2018mosculp} produces a sculpture representing a moving object or person (e.g., a dancer).
Motion Structures~\cite{reyes2013motion} turns video frames (e.g., Game of Thrones teaser) into 3D printed sculptures.

\subsection{Other Datasets}
\label{subsec:other}

In \numOtherData of our reviewed papers and projects, we found datasets that did not fit in the aforementioned categories. An example of these \textit{other} datasets is motion, action, and movement, which can result in interesting physical patterns rich in details. Motus Forma~\cite{Allen2016Motus} captures 10 hours of people's movement in a lobby space, with more than 1300 motion paths. By attaching sensors to the back of crochet hooks and combining the data into 3D coordinates via a Processing script, Nissen and Bowers designed path-like patterns to capture hand movements of crochet practitioners with varied skill levels~\cite{nissen2015data}. With the goal of understanding various activities within a FabLab environment, \textit{ Cairn}~\cite{gourlet2017cairn} is a collaborative sculpture with various laser cut pieces. \autoref{table:other datasets} summarizes different types of datasets under the \textit{other} category in this survey, along with their corresponding works. 

\begin{table}[h!]
\centering
\begin{tabular} {  m{3cm}  m{4.5cm}  } 
 
 \hline
 
\centering \textbf{Entry} & \textbf{Dataset} \\
 
\hline\hline
 
\centering \cellcolor[HTML]{F6F6F6} \cite{Allen2016Motus,nissen2015data,gourlet2017cairn,Lozano2004Array,kazi2016chronofab, BMW2008Kinetic,rowe2012within} & \cellcolor[HTML]{F6F6F6} Motion, Action, and Movement  \\
 \hline
\centering\cite{schuller2016computational,zhang2015computational,tang2019computational,li2018construction,tena2013fabricating,thun2012soundspheres} & 3D Patterns and 3D Objects \\
\hline
\centering \cellcolor[HTML]{F6F6F6} \cite{patel20173d,miyashita2016zoematrope,tymms2018quantitative,dumas2015example} & \cellcolor[HTML]{F6F6F6} Texture and Material \\
\hline
\centering \cite{rodighiero2018printing,khot2016fantibles,nissen2015data,starrett2018data,isketch2016Podium,Chen2014Xpose} & Social Media, Network, and Society \\
\hline
\centering \cellcolor[HTML]{F6F6F6} \cite{Gungor2011Blip} & \cellcolor[HTML]{F6F6F6} Travel Data \\
\hline
\centering\cite{Goods2010eCloudAirFIELD,moere2009physical} & Aviation Data \\
\hline
\centering \cellcolor[HTML]{F6F6F6} \cite{Schneider2012Lego,schneider2015visualisation,swaminathan2014supporting} & \cellcolor[HTML]{F6F6F6} PhD Studies \\
\hline
\centering\cite{marcus2014centennial} & Different Degree Type Offered \\
\hline
\centering \cellcolor[HTML]{F6F6F6} \cite{madura20153d} & \cellcolor[HTML]{F6F6F6} Astronomy \\
\hline
\centering\cite{Something2014Tree,epler2012grand} & Public Opinion \\
\hline
\centering \cellcolor[HTML]{F6F6F6} \cite{Heinicker2015Necklace,koutsomichalis2018objektivisering,goni2016deletion,geiger2014datenreise,katsumoto2018robotype} & \cellcolor[HTML]{F6F6F6} Words, Terms, and Text \\
\hline
\centering\cite{Kison2009Pulse,moere2009physical} & Emotions and Relationship Status \\
\hline
\centering \cellcolor[HTML]{F6F6F6} \cite{LeGoc2018dynamic} & \cellcolor[HTML]{F6F6F6} Tourist Peak Periods \\
\hline
\centering\cite{Shumay2016FizViz,Keller2009DataMorphose} & Website Traffic \\
\hline
\centering \cellcolor[HTML]{F6F6F6} \cite{le2016zooids,swaminathan2016linespace} & \cellcolor[HTML]{F6F6F6} Freehand Drawing \\
\hline
\centering\cite{holstius2004infotropism} & Amount of Trash and Recyclables \\
\hline
\centering \cellcolor[HTML]{F6F6F6} \cite{Barry2012Saturation} & \cellcolor[HTML]{F6F6F6} FM Radio Spectrum \\
\hline
\centering\cite{gwilt2012enhancing} & Package Openability \\
\hline
\centering \cellcolor[HTML]{F6F6F6} \cite{zoran2018digital} & \cellcolor[HTML]{F6F6F6} Taste Structures \\
\hline

\end{tabular}
\caption{Other datasets used for physicalizations.}
\label{table:other datasets}
\end{table}

\section{Design and Physical Rendering Approaches}
\label{sec:rendering}

In this section, we discuss methods used to make a visual presentation and bring it into the physical world. Our goal is to discuss various approaches used for design and physical rendering, using different digital design and fabrication tools. Based on the reviewed works in our corpus, a typical process planning for the physical rendering process consists of design sketching, making accurate 3D representations of the physicalization design, AKA 3D modelling, physical prototyping, modifying the design (i.e., iterative design), final fabrication, and conducting studies (see Section \ref{sec:discussion} for iterative design and user studies).


\subsection{Design of Physicalizations}
\label{Design}

In this report, physicalization design is the stage of making the abstract visual form and the final visual presentation, i.e., visual mapping and presentation mapping as introduced in \cite{jansen2013interaction}. While this step is full of opportunities, it also introduces several challenges for visualization designers who have always considered cognition and perception for their on-screen or paper-based designs. When working in physicalizations, visualization designers should consider perception and experience of physical environments, materiality, cultural symbolism, and spatial relationships. Many of these challenges have been explored and practiced for many years in the fields of industrial design and architecture. As a result, investigating the design principles and steps architects and industrial designers take can be quite helpful for the design of physicalizations as well. Sosa et al. have introduced four design principles inherited from industrial design that can be applied to physicalizations\cite{sosa2018data}. They encourage physicalization designers to treat data as a new type of material to design with, design for (re)interpretation of the target data, design for cognitive and emotional engagement with target data, and design to give people the opportunity to use the data to rethink. Cull and Willet propose the concept of \say{data tectonics} to describe the holistic nature of designing physicalizations\cite{hull2018data}. As an integrative theory, tectonics in architecture examines \say{the interwoven relationship between space, function, structure, context, symbolism, representation and construction.}\cite{schwartz2016introducing}. Data tectonics defines \say{the
relationship between context, data, visual representation, materiality, fabrication and interactions of a data representation} and suggests that physicalization designers borrow from the approaches used by architects for many years (e.g., design sketching, diagramming, and making scale models)\cite{hull2017building,hull2018data}.

\begin{figure*}[t]
  \centering
   \includegraphics[bb=0 0 1880 573, width=\textwidth]{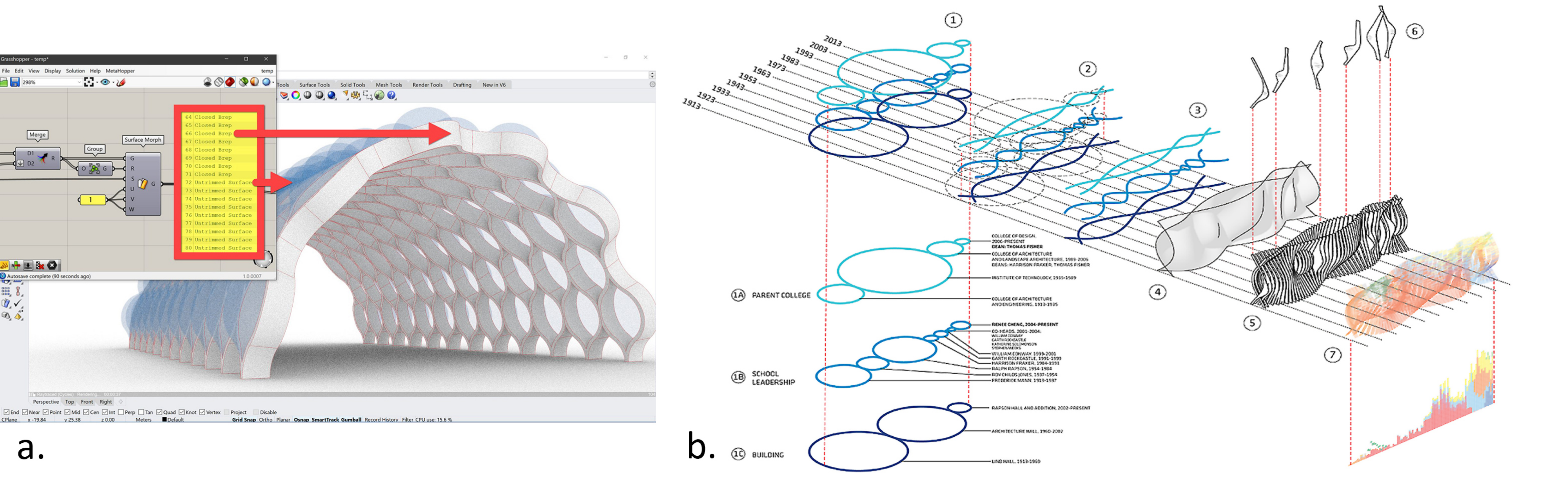}
   \caption{(a) An example of a parametric design generated by Grasshopper. The visual programming interface of Grasshopper, as well as its data list, is shown on the left. Image taken from\cite{sondermann2009Parametric}, (b) A diagram showing parametric design stages of generating form from the original data, using Grasshopper. Image taken from \cite{marcus2014centennial} }
           \label{fig:Grasshopper}
\end{figure*}

\subsection{Digital Design}
\label{subsec:DigitalDesign}

Design is the first stage of the rendering process that applies visual mapping transformation to data and gives it an initial visual form. The introduction of CAD and its ability to deal with more complex geometrical problems \cite{khabazi2010generative} has made digital design a popular approach for physicalization. CAD, as an umbrella term, covers a vast array of tools that produce different results such as 2D drawings and 3D models. CAD data has the great option of transferability into other software platforms to control the appearance and other formal characteristics of physicalizations \cite{dunn2012digital}.

\subsubsection{2D CAD}

For various physicalization scenarios, a 2D drawing needs to be made in CAD. This 2D drawing can be either a continuous path (vector) or a discrete path (raster), such as a series of images. It is the output of the processing pipeline of the fabrication technique that should be considered for making decisions about creating vector or raster designs (e.g., cutting lines vs. engraving images in laser cutting).

2D CAD is usually used for preparing outlines and contour lines to be used for laser cutting, such as the pieces making the \textit{Trend in Water Use} sculpture\cite{Adrien2011data} or tokens representing people's activities in FabLabs in \textit{Cairn}\cite{gourlet2017cairn}. One of the frequently used CAD software to make vector 2D drawings for physicalization purposes is Adobe Illustrator. For instance, H{\"a}kkil{\"a} and Virtanen have translated the collected sleep data from an Oura ring to 2D charts and 2D paths for laser cutting, using Illustrator \cite{hakkila2016aesthetic}. There may be some design cases for physicalizations that hand-drawn sketches of paths should be translated into vector data. In such cases, Illustrator can be used to trace over scanned hand-drawn paths, such as the 1306 individual paths showing the movement of people in Motus Forma \cite{Allen2016Motus}. Outputs from programming-based CAD designs (see Section \ref{subsubsec:3DCAD}) can be exported to Illustrator to make laser-cut ready vector files. Such files include various line types, based on the defined paths (e.g., cutting once or twice) and actions (e.g., cutting or engraving) for laser cutters (see Section \ref{subsub:subtractive} for more details). An example of such application for Illustrator is \textit{Blip}, which has transformed a year of travel into data sculptures\cite{Gungor2011Blip}. 

Vector paths created by 2D CAD software can also be used as part of the modelling process in any 3D CAD platform to make volumetric designs and generate suitable files for fabrication. In the following section, we will cover various scenarios for 3D CAD modelling that can be used for the design of physicalizations.


\subsubsection{3D Modelling}
\label{subsubsec:3DCAD}

To model 3D objects that can be fabricated, three primary representations are usually used: polygonal meshes, Non‐Uniform Rational B‐Splines (NURBS), and constructive solid geometry (CSG). 

Polygonal meshes provide a discrete representation in which an object is represented by a set of polygonal facets indicating the connectivity of the shape along with a set of vertices with $(x,y,z)$ coordinates providing the geometry. Due to the simplicity and effectiveness of this representation, meshes are industry standards and are included in many 3D modeling software programs including Maya\cite{Maya} and Blender\cite{Blender} and they have been also used for the sake of physicalization (e.g., \cite{barrass2011phsyical,barrass2012digital}).

\begin{figure}
  \centering
   \includegraphics[bb=0 0 2173 815, width=\linewidth]{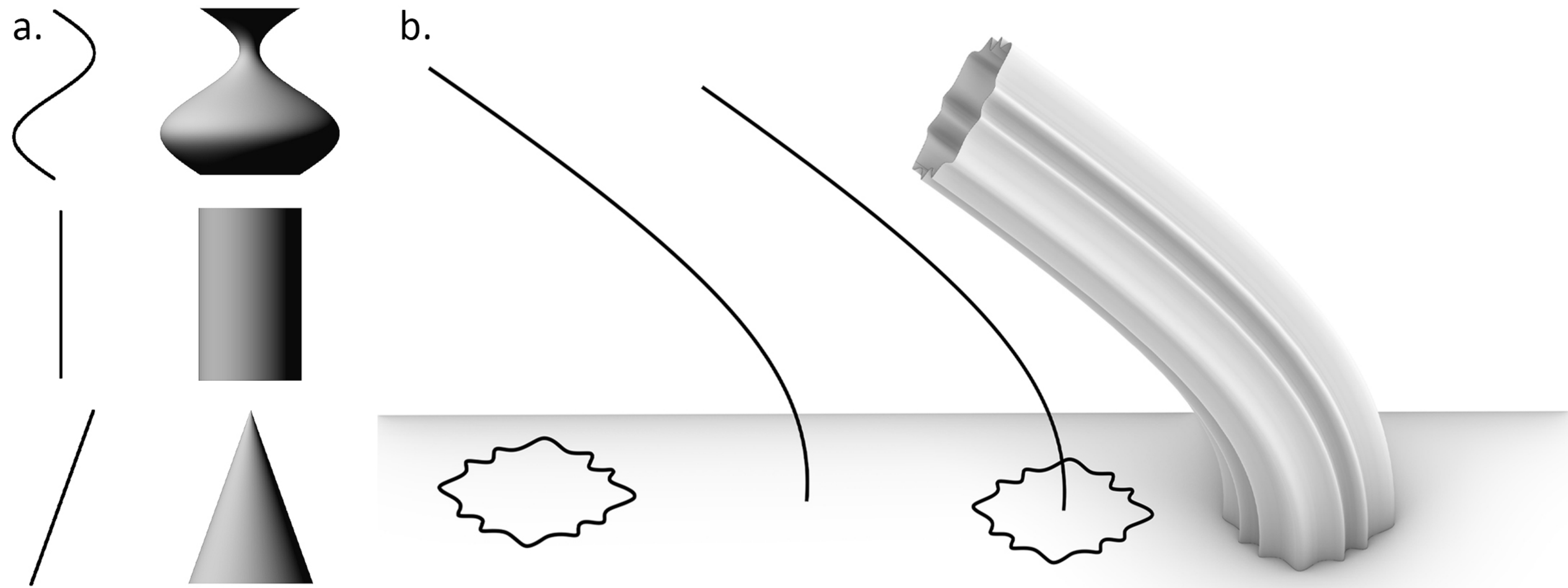}
   \caption{Examples of NURBS surfaces generated by attaching several NURBS patches: (a) Surface of revolution, (b) Sweep surface. Images taken from \cite{samavati2013modelling}.}
           \label{fig:NURBS}
\end{figure}

To offer designers a higher degree of control on the form, digital modelling programs also utilize continuous curve and surface representations in which a model can be modified by a set of control points. NURBS are powerful representations in this setting as control points can attain different weights to push or pull a curve or surface; a property that other representations such as B-Splines do not have and therefore they are limited in producing many simple and complicated shapes including a circle.
NURBS can be directly used to create curves and surface patches. It is also possible to make a 3D shape by attaching several NURBS patches or generate a 3D surface from a profile curve using techniques such as the surface of revolution or sweep surfaces (see \autoref{fig:NURBS}).
Due to these powerful features, NURBS is very popular in physicalization \cite{Shumay2016FizViz,katsumoto2018robotype,hakkila2016aesthetic}.

Although the curves and surfaces produced by NURBS provide a high degree of flexibility via control points and weights \cite{dunn2012digital}, some designers, especially for designing CAD shapes, prefer to use CSG since it provides sharp and accurate final results. In CSG, a shape is produced by applying several operations (e.g., union, intersection, difference, etc) on simple shapes such as spheres and cylinders to produce an accurate final object.
CSG has been also used for physicalizations such as the customized Lego-Bricks provided by Schneider \cite{Schneider2012Lego}.

 

In some physicalization scenarios, the 3D models are designed, developed, and made ready for fabrication by only using various CAD software packages and their features and functionality. 
Examples of such software programs are Maya \cite{Maya}, 3D Studio Max \cite{3DMax}, Blender \cite{Blender}, and Rhino \cite{Rhino}.
For instance, NURBS provided in Rhinoceros®, AKA Rhino, has been used to physicalize various models (e.g.,\cite{Shumay2016FizViz,katsumoto2018robotype,hakkila2016aesthetic}). We refer to such modelling as \textbf{CAD} in our taxonomy (see \autoref{tab:Taxonomy1} and \autoref{tab:Taxonomy2}). 

To ease the process of modelling, many software programs provide a Visual Programming interface, where users connect a series of functional blocks into a sequence of actions. The only required \say{syntax} in such method is that each block should receive the appropriate data types as its input. 
Such solution is referred to as \emph{parametric design} \cite{dunn2012digital}. Note that this term is different from parametric representation, such as NURBS and B-Splines, in which shapes are defined by benefiting from a parameter space.
As a rigorous rule-based system, parametric design involves precise, step-by-step techniques that make multiple options based on a set of rules, inputs, and values specified by designers \cite{dunn2012digital, wassim2013parametric}. 
Grasshopper®, a visual programming plug‐in designed for Rhino®, is one of those mediums that has a visual interface and its components can provide, manipulate, and modify data, as well as draw and modify objects (see \autoref{fig:Grasshopper}). 
Grasshopper has been extensively used to produce physicalization techniques \cite{Aweida2013RobotArranges,velikov2014nervous,hosseini2019data,marcus2014centennial,Huang2017moire}. This type of design is called \textbf{Parametric Design} in our taxonomy (see \autoref{tab:Taxonomy1} and \autoref{tab:Taxonomy2}).

User interfaces for 3D modeling commonly follow the WIMP (Windows, Icons, Menus, Pointer) paradigm \cite{JS11}. Sketch-based interface is  considered as an alternative paradigm for 3D modeling \cite{OSSJ08}. In this approach, 2D hand-drawn sketches are used in the modeling process,  from model creation to editing and augmenting the initial model in an iterative manner \cite{OS10,OSSJ05}. 

Extra development and customization sometimes have been employed as pre-processing, post-processing or in the form of scripting to prepare data or add necessary functionality.
For example, Processing\cite{Processing} has been used to produce line graphs of voter approval rate data, available on the Internet, before making 3D shapes for fabrication \cite{epler2012grand}. To physicalize geospatial datasets, the coarse geometry of the Earth has been first extracted from a Digital Earth platform and then Rhino is used to develop the forms, design data attachment details, and make the pieces fabrication-ready \cite{djavaherpour2017physical,moorman2020geospatial}. Scripting has been performed to make 3D models and hinges for producing a mathematical puzzle benefiting from CSG operations available in Blender \cite{li2018construction}.
Parametric design platforms (e.g., Grasshopper) are also compatible with script‐based programming languages such as Python to make custom algorithms for the design of physicalizations. For instance, Hosseini et al. \cite{hosseini2020portal} have used Grasshopper and custom Python scripting to build Portal. In our taxonomy table (see \autoref{tab:Taxonomy1} and \autoref{tab:Taxonomy2}), we have referred to such design approach as \textit{Hybrid}.


There are many cases in the design of physicalizations where off-the-shelf CAD software, and even parametric or hybrid design approaches, are not able to handle the complexity of the process of transforming data into a model. In such cases, physicalization designers make their own programs via available programming languages and libraries (e.g., C++ and OpenGL). Many different programming languages have been used for physicalization, among which Processing, an open-source Java-based language developed for designers, is the most popular. The Processing community has written more than a hundred libraries to facilitate computer vision, data visualization, 3D file exporting, and programming electronics \cite{Processing}. Depending on the community, other programming languages such as Python, Java, or C++ have been also utilized to make a customized modelling program.
Physicalizations for which a standalone program has been produced include Landscaper\cite{allahverdi2018landscaper}, works to add textures on 3D prints \cite{schuller2016computational,zhang2015computational,mahdavi2015coverit}, make water color paintings \cite{panotopoulou2018watercolor}, etc.


\begin{figure*}[t]
    \centering
    \includegraphics[bb=0 0 4521 815, width=\textwidth]{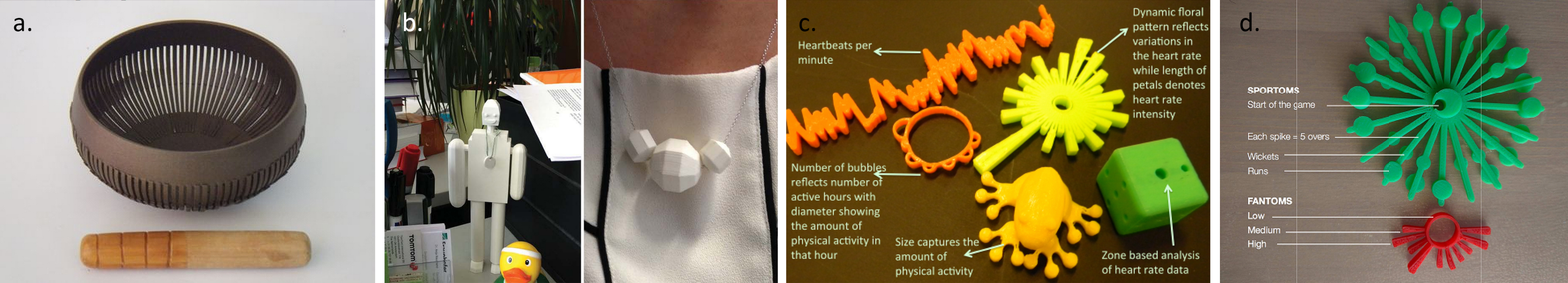}
    \caption{FDM 3D printing has been used to produce physicalizations of health, personal, and sports data: (a) The Hypertension Singing Bowl is a stainless steel 3D printed sonification that has transformed blood pressure data to a sculpture that rings. Image taken from\cite{barrass2014acoustic}, (b) A figure and a necklace sculpture physicalizing running activity. Image taken from\cite{stusak2014activity}, (c) Five different representations of physical activity, each focusing on one specific aspect. Image taken from\cite{khot2014understanding}, (d) Fantibles is a physicalization summarizing sports data and user excitement. Image taken from\cite{khot2016fantibles}.}

    \label{fig:Personal}
\end{figure*}


\subsection{Digital Fabrication}
\label{subsec:fabrication}

Fabrication makes the visual presentation perceivable by bringing it into existence in the physical world \cite{jansen2013interaction}. In digital fabrication, computer-controlled manufacturing machines receive digital models to build 2D or 3D objects \cite{swaminathan2014supporting}. There are two main approaches to digital fabrication: 1) Subtractive techniques (e.g., laser cutting and CNC milling) that cut away or remove material, 2) Additive techniques (e.g., 3D printing) that build up material layer‐by‐layer \cite{swaminathan2014supporting}. However, some references categorize digital fabrication techniques as cutting, subtractive, additive, and formative \cite{kolarevic2003digital,dunn2012digital}. Fabrication techniques can also be categorized as 2D or 3D. An example of the former is laser cutting that operates on flat sheets of material and examples of the latter are CNC mills and 3D printers that produce 3D solid objects.

The primary challenge when fabricating objects is to ensure that they embody the necessary physical properties including a) Cost, b) Manufacturability or Ease of Fabrication, c) Assembly and Fit, d) Statics (Balance, Stability, and Strength), and e) Fabrication-specific Effects \cite{swaminathan2014supporting,Hullin2013Computational}. Regarding balance, stability, and strength, some software packages facilitate the design iteration through simulation, such as AutoDesk® Inventor and heatmap stress visualization applications. We will cover these challenges in more details in Section \ref{sec:discussion}. 

Table \ref{table:digital fabrication} summarizes different attributes related to each category of digital fabrication techniques. This section is continued by introducing various tools in each of the digital fabrication categories.

\begin{table}[h!]
\centering

\begin{tabular} {  m{1.5cm}  m{6.1cm}  } 
 
 \hline
 
 \centering \textbf{Technique} & \textbf{Attributes} \\
 
 \hline\hline
 
 \centering \cellcolor[HTML]{F6F6F6} Cutting  & \cellcolor[HTML]{F6F6F6}
 \begin{itemize}
 \bigskip
     \item Easily accessible,
     \item Makes shaped 2D elements from sheet materials
     \item Cutting Methods: Laser, Water Jet, Plasma Arc
 \end{itemize} \\
 
 
\centering Subtractive & \begin{itemize}
 \bigskip
    \item Takes material from an existing solid volume and creates the desired shape,
    \item Axially, surface, or volume‐constrained cutting heads
    \item Advantages:
    \begin{enumerate}
        \item Larger component size,
        \item Wider range of material selection,
        \item More precise fabrication,
    \end{enumerate}
\end{itemize} \\


\centering \cellcolor[HTML]{F6F6F6} Additive & \cellcolor[HTML]{F6F6F6} \begin{itemize}
 \bigskip
    \item Converts CAD to a series of 2D layers, i.e., layer-by-layer fabrication (AKA rapid prototyping)
    \item Advantages:
    \begin{enumerate}
        \item Direct \say{file to fabrication} process,
        \item Fabricates complex forms,
        \item Non‐expert use,
    \end{enumerate}
    \item Disadvantages: limited size, limited range of materials, lengthy production times
    \item Examples include: 3D printing techniques (Fused Deposition Modelling (FDM), Stereolithography (SLA), Direct Metal Laser Sintering (DMLS), Selective Laser Sintering (SLS), Selective Laser Melting (SLM), Electron Beam Melting (EBM)), knitting machines
\end{itemize} \\


\centering Formative & \begin{itemize}
 \bigskip
    \item Uses mechanical force, heat, and steam to reshape
    \item Can be axially or surface constrained
    \item Examples include: vacuum forming, thermoforming (after 3D printing)
\end{itemize} \\

\hline

\end{tabular}

\caption{An overview of digital fabrication tools and techniques.}
\label{table:digital fabrication}
\end{table}

\subsubsection{2D Printing}

A trivial technique to bring patterns, designs, and visualizations into the physical world is traditional (2D) printing. In 2D printing, key parameters are the print resolution and the printer gamut defined by the inks or toners employed\cite{Hullin2013Computational}. When used in creative setups, such as the installation made by Kyriaki Goni\cite{goni2016deletion} or the re-invention of Volvelles in the work of Stoppel and Bruckner\cite{stoppel2016vol}, 2D printing can be a powerful tool to make an engaging physicalization that is well capable of conveying the message of its target data. Moreover, using 2D printing in physicalization is a method that supports the possibility of having various colours
and to overcome the limitations of affordable off-the-shelf FDM 3D printing (see Section \ref{subsubsec:additive}).

\subsubsection{Additive Techniques}
\label{subsubsec:additive}

The general concept in additive manufacturing is to build objects layer-by-layer from a small number of basis materials \cite{Hullin2013Computational}. One of the most common tools that digitally fabricate objects with an additive approach are 3D printers. Over the past decade, 3D printers have become more accessible to the consumer market with low maintenance and operating costs. Moreover, the possibility of making complex objects by using 3D printers have made them a common choice for making prototypes or final physicalizations. Different types of 3D printers exist that all build objects on a layer-by-layer basis, but some \textit{locally deposit} material and some \textit{solidify} material within a non-solid substance \cite{livesu20173d}. We have \numPrint works using various methods of 3D printing.  

\begin{figure}
  \centering
   \includegraphics[bb=0 0 3198 815, width=\linewidth]{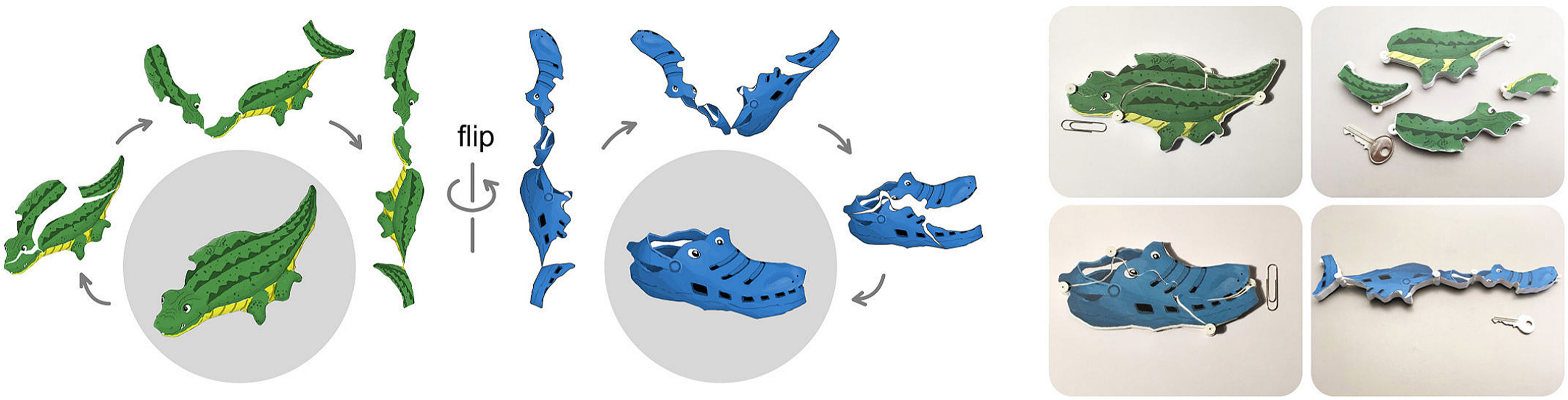}
   \caption{3D printing has been used to make mathematical puzzles. The crocodile and the Crocs shoe can be inverted inside-out and transformed into each other in a fully automatic manner. Image taken from \cite{li2018construction}.}
   
   \label{fig:MathPuzzle}
\end{figure}

\noindent\textbf{FDM 3D Printing.} One of the most accessible and affordable 3D printers are Fuse Deposition Modelling (FDM) printers that make 3D objects layer‐by‐layer through heating and extruding thermoplastic or wax filaments \cite{zhang20163d}. 
FDM 3D printing has a long history in physicalizing complicated shapes such as macromolecular assembly \cite{bailey1998use}. Until now, many physicalizations have been produced via FDM printing for different applications such as education \cite{brown2012viztouch,moorman2020geospatial,kane2014tracking}, project management \cite{Schneider2012Lego}, producing geological artifacts \cite{hosseini2019data,McCune2013Maps,allahverdi2018landscaper,luo2012chopper,dadkhahfard2018area,kirshenbaum2020data}, visualizing health, sport, or other personal data \cite{barrass2014acoustic,khot2014understanding,khot2020shelfie,khot2016fantibles,stusak2014activity,Heinicker2015Necklace} (see \autoref{fig:Personal}), generating 3D models from text \cite{koutsomichalis2018objektivisering}, mathematical puzzles \cite{li2018construction} (see \autoref{fig:MathPuzzle}), environmental data \cite{Whitelaw2009Weather}, astrophysical \cite{madura20153d} and statistical data \cite{Mccune2016Housing,gwilt2012enhancing}, or even thoughts \cite{geurts2018imagining}.

\begin{figure}
  \centering
   \includegraphics[bb=0 0 2743 815, width=\linewidth]{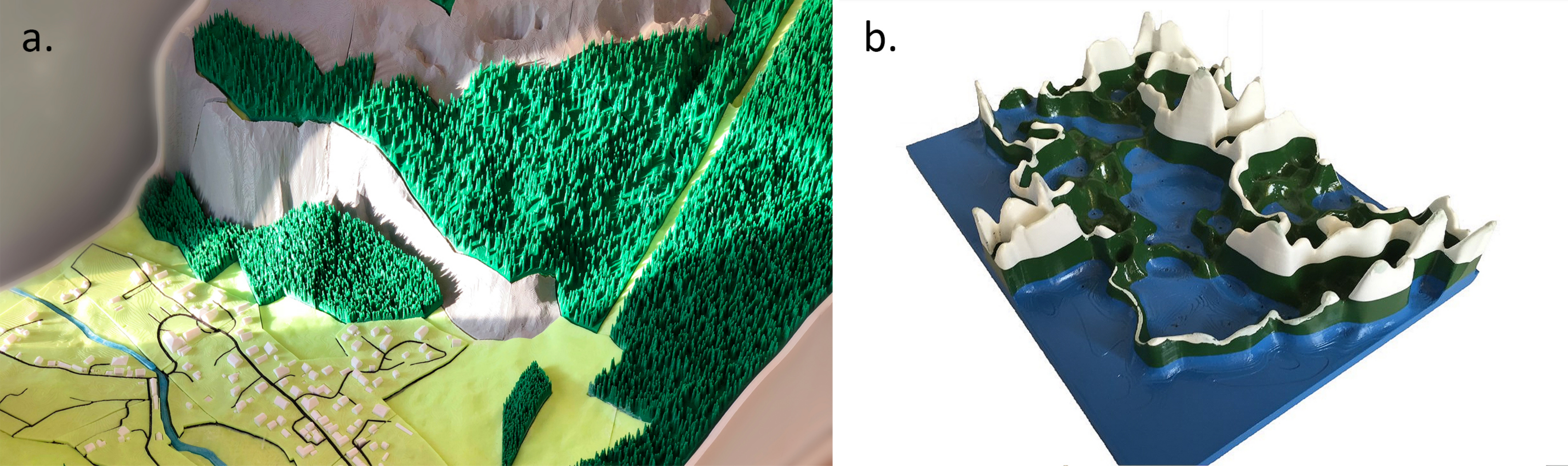}
   \caption{One solution to overcome the lack of colour in 3D printing is fabricating the physicalization in discrete pieces, each with a different filament colour, and assemble them. Image (a) taken from \cite{allahverdi2018landscaper} and image (b) taken from \cite{thrun2016visualization}.}
   
   \label{fig:FDMColour}
\end{figure}

However, since material in this approach needs to be deposited on top of an exiting layer, FDM printing relies heavily on support structures that need to be removed after fabrication.
Therefore, to fabricate delicate structures, creative solutions such as layer supports are designed \cite{ang2019physicalizing}.
FDM printing also suffers from a limited building volume, which is by average $20^3 cm^3$. Therefore, breaking a large model into printable volumes have been employed \cite{allahverdi2018landscaper,luo2012chopper}.
In addition, the results of FDM are usually limited in terms of number of colors, therefore innovative solutions have been proposed to overcome these challenges. To resolve this problem, geological features with different properties are printed in different but limited colors \cite{thrun2016visualization,allahverdi2018landscaper} (see \autoref{fig:FDMColour}). Projectors have been also used to visualize data on a base model \cite{dadkhahfard2018area}.

\noindent\textbf{Layer Solidification 3D Printing.} Layer solidification is a 3D printing process in which the top (or bottom) surface of the object is solidified from a non-solid material, such as liquid or powder, within a tank. This process is executed by vat photopolymerization (e.g., stereolithography or SLA 3D printers), powder bed fusion (e.g., Selective Laser Sintering or SLS 3D printers, binder jetting (e.g., plaster powder binding), and sheet lamination (e.g., paper layering–cutting) \cite{livesu20173d}.
Although this type of 3D printing still has some limitations such as the size of the resulting products, it does not need to print additional support structures and therefore the final products have better surface quality.
As a result, it is possible to produce delicate physicalizations using this technique such as data sculptures in the form of a tree \cite{Something2014Tree} that is difficult to produce by an FDM printer. 
However, 3D printing by this technique is usually more expensive than FDM.

\begin{figure}
  \centering
   \includegraphics[bb=0 0 4197 815, width=\linewidth]{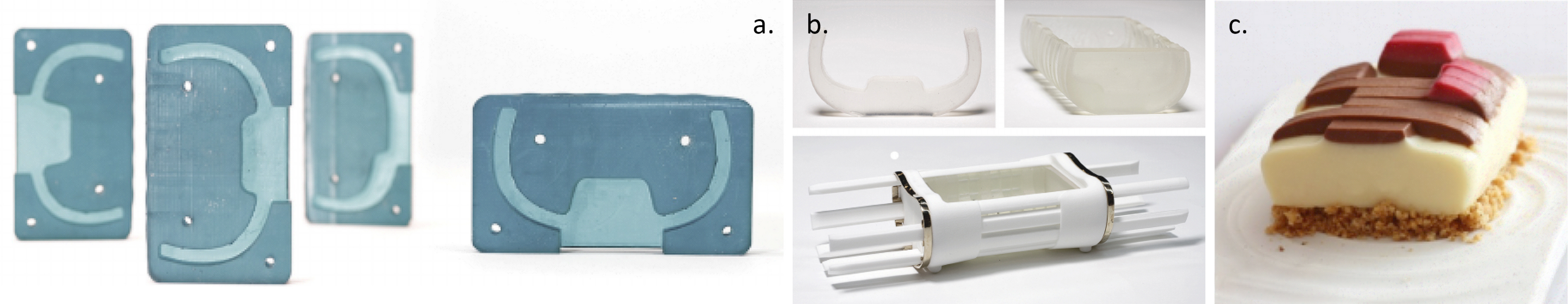}
   \caption{SLA 3D printing has been used to make plastic molds (a) for silicone casting (b) to bake cakes representing various taste structures (c). Images taken from\cite{zoran2018digital}.}
   
   \label{fig:cooking}
\end{figure}

There are \numSLS projects that benefit from this type of 3D printing for different applications and in different area such as biology \cite{gillet2005tangible}, medical \cite{nadeau2000visualizing}, statistics \cite{Kauffman2013NYC}, artistic furniture \cite{zhao2016printed} and data sculptures \cite{Something2014Tree,starrett2018data}, reservoir field exploration \cite{nittala2015planwell}, physicalizing videos \cite{zhang2018mosculp} and sound \cite{barrass2012digital}, even cooking molds \cite{zoran2018digital} (see \autoref{fig:cooking}).

\begin{figure}
  \centering
   \includegraphics[bb=0 0 5619 815, width=\linewidth]{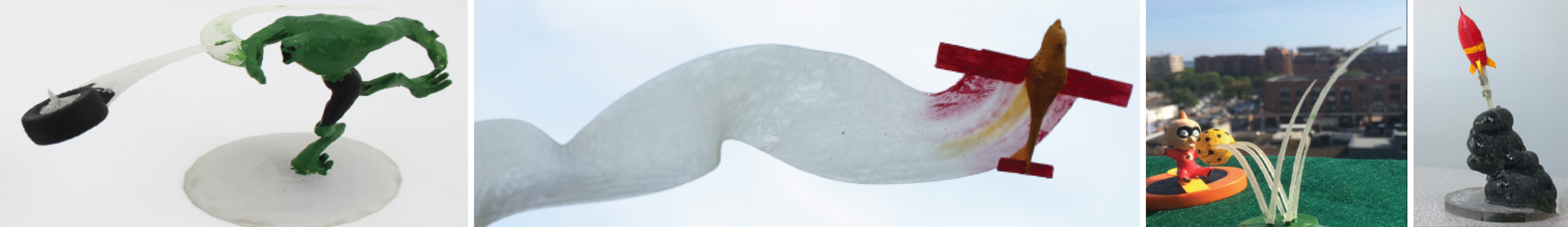}
   \caption{Chronofab uses Stratys 6-colour 3D printers to fabricate motion. Images taken from\cite{kazi2016chronofab}.}
   
   \label{fig:chronofab}
\end{figure}

\noindent\textbf{3D Colour Printers.} Since colors play an important role in an understandable visualization, color 3D printers (e.g., ZCorporation multi-colour 3D printer) have been  used to produce geological physicalizations \cite{rase2011creating}. Stratys 6-colour 3D printer has been used to fabricate beautiful voxelized data \cite{bader2018making} or motion \cite{kazi2016chronofab} (see \autoref{fig:chronofab}) and Connex3 500 has been used to physicalize a variety of creative objects that can handle deformation or attain specific textures\cite{patel20173d}.

\begin{figure*}[t]
    \centering
    \includegraphics[bb=0 0 4644 769, width=\textwidth]{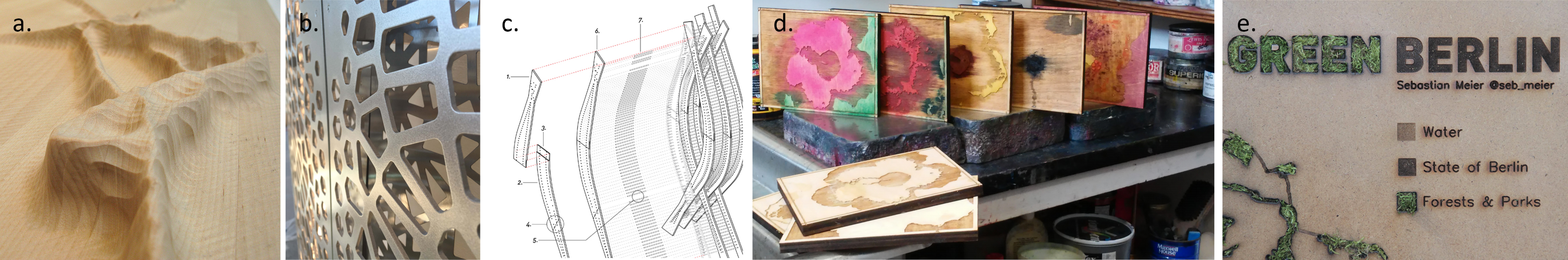}
    \caption{Subtractive techniques, such as CNC and laser cutting have been used in many physicalizations: (a) Sculpture carved out of a block of wood by a CNC milling machine, showing wind directions. Image taken from\cite{Knapek2012Windcuts}, (b) Metal panel cut by a CNC machine to show IBM sales in the form of a data spatialization. Image taken from\cite{Huang2017moire}, (c) An example of shop drawings and technical assembly diagrams for physicalizations that are fabricated in pieces and require assembly. Images taken from\cite{marcus2014centennial}, (d) Various woodblocks prepared by laser cutting to make a watercolour painting of a flower. Image taken from\cite{panotopoulou2018watercolor}, (e) Carved out parts of an MDF sheet are filled with moss to make a living map of forests. Image taken from\cite{Meier2017Green}.}

    \label{fig:CNC}
\end{figure*}

\subsubsection{Subtractive Techniques}
\label{subsub:subtractive}

Subtractive manufacturing techniques are on the opposite side of 3D printing. In other words, rather than incrementally building up a model, subtractive techniques gradually remove material from an unmachined part by using a sharp cutting tool\cite{livesu20173d}. Milling is the most versatile subtractive technique with a large variety of materials available for it.

\noindent\textbf{CNC.} Computer Numerical Control (CNC) is one of the most commonly applied methods of digital fabrication\cite{aitcheson20053axis,kolarevic2003digital,dunn2012digital}, used by \numCNC projects in our corpus. CNC has the potential to fabricate double-curved and developable surfaces \cite{kolarevic2003digital,aitcheson20053axis}. In CNC milling, stepper motors control the movement of the individual axes of tool movement. Two types of artifacts are common with CNC machines: step artifacts and tool path artifacts that leave tiny grooves on the final model \cite{livesu20173d,mahdavi2020VDAC}. This is something that physicalization designers need to consider when planning on the digital fabrication tools, as some artifacts may be misunderstood as data.

CNC milling can be used to make moulds for the next steps of the fabrication process (e.g., \cite{tabrizian2017tangible}), or to make the final physicalization (e.g., \cite{marcus2014centennial,Huang2017moire,Jarman2014Cosmos,priestnall2012projection,Knapek2012Windcuts,Hush2014MadebyNum,swaminathan2014supporting}) (see \autoref{fig:CNC}a,b). In fabrication cases that a huge number of pieces have to be milled separately and assembled (see section \ref{subsubsec:assembly} for more information on the assembly process), detailed shop drawings have to be produced to support a consistent tolerance throughout the assembly process (e.g., \cite{marcus2014centennial}) (see \autoref{fig:CNC}c).

\noindent\textbf{Cutting Techniques.} Cutting techniques can be considered as a sub-category of subtractive methods. One of the most popular cutting methods is laser cutting (used by \numLaser works in our corpus), mostly due to its speed, efficiency, and its ability to cut a wide range of materials \cite{swaminathan2014supporting}. Laser cutting have been used by the physicalization community to make 3D bar charts (e.g., \cite{jansen2013evaluating}), stacked scatter plots (e.g., \cite{Stusak2013Layered,Gungor2011Blip,Dwyer2005Time-evolving}), data sculptures (e.g., \cite{gourlet2017cairn,Madsen1995Early,Perovich2014Data}), and active physicalization and spatialization (e.g., \cite{velikov2014nervous,houben2016physikit}. In order to support the making of active physicalizations solely by using laser cutters, Polysurface\cite{everitt2017polysurface} has introduced a novel fabrication method. This proposed method fabricates elastically deformable sheets out of a single sheet of polypropylene, attached to spandex for fluidity. Since laser cutters cut out outlines and silhouettes from sheets of material, the direct result of their fabrication process is a 2D object extruded by the material thickness. To overcome this issue, one common solution is cutting several pieces and stacking them up to form a contoured 3D object (e.g.,\cite{fens2014personal,Stusak2013Layered}. Such approach has risen some sustainability challenges, both in terms of time and material, that we will discuss in Section \ref{sec:discussion}.

Laser cutters are capable of etching the surface of different materials and provide engraving. This is quite handy when details need to be added to physicalizations, such as some information about charts axes. Moreover, this ability of engraving, combined with material properties, can be used to make the whole physicalization, such as producing woodblock printing of watercolour paintings\cite{panotopoulou2018watercolor} (see \autoref{fig:CNC}c). Such removal of the material also provides the opportunity to fill the holes with different materials and make novel physicalizations. An example of such approach is Green Berlin\cite{Meier2017Green}, which has made a living map of forests and parks in Berlin by filling the cut-away parts of wood with moss (see \autoref{fig:CNC}d).

Another well-known cutting technique is waterjet cutting that is capable of cutting a wide variety of materials by using a high-pressure jet of water and an abrasive material. In contrast to laser cutting, which only requires the material for the cutting process, waterjet consumes massive volumes of water and abrasive material, non of which are recyclable (see Section \ref{sec:discussion}).

\subsubsection{Formative Techniques}

\begin{figure}
  \centering
   \includegraphics[bb=0 0 1277 516, width=\linewidth]{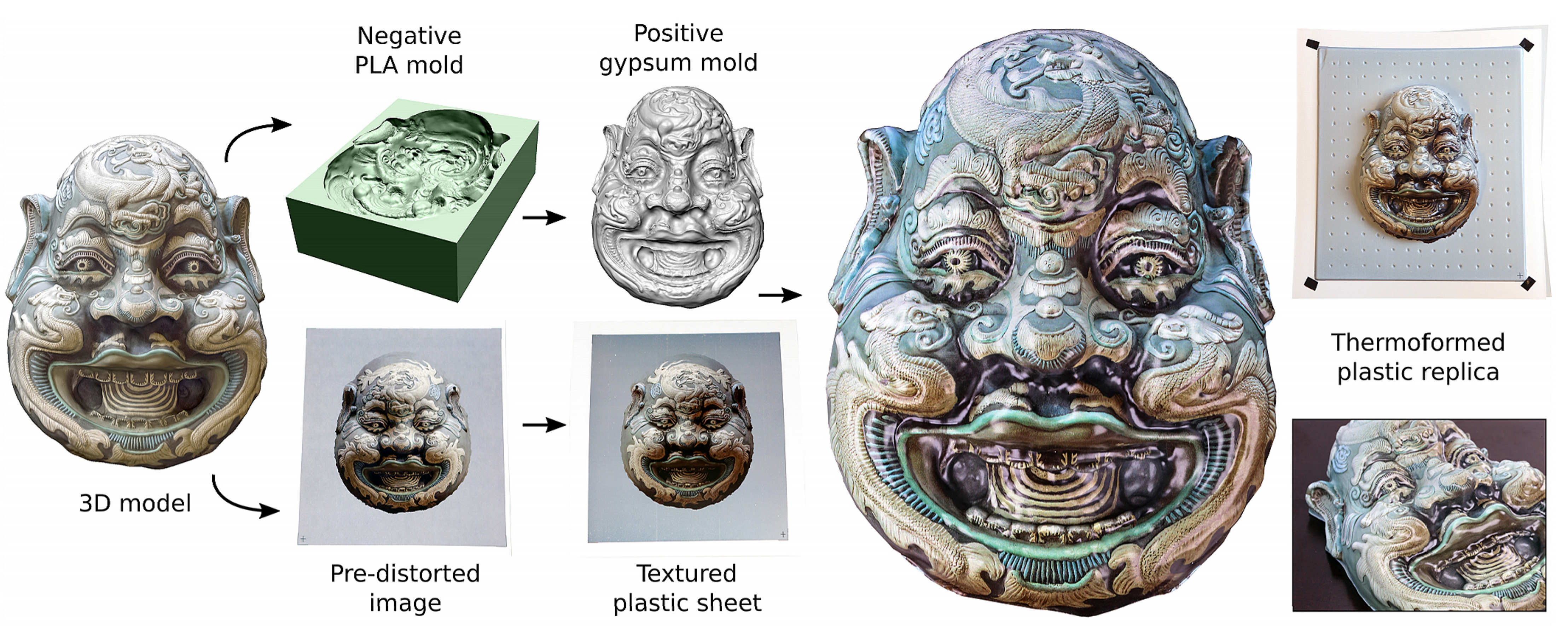}
   \caption{Computational Thermoforming is an advance method to add colour and texture to physicalizations. Image taken from\cite{schuller2016computational}.}
   
   \label{fig:thermoforming}
\end{figure}

Formative fabrication processes utilize mechanical forces to reshape or deform materials into the required shape. Examples of formative approaches are \textit{vacuum moulding} (i.e., heating a thermoplastic sheet of material until it becomes malleable and then sucking it on a shape using vacuum pressure) and \textit{thermoforming} (i.e., heating a sheet of plastic material until it becomes malleable and then forming the sheet onto a forming core shape). Computational Thermoforming\cite{schuller2016computational} introduces a novel method for the fabrication of textured 3D models. This approach is meant to be used for customized, unique objects, which makes it a useful solution to support colour and texture for physical rendering of physicalizations. \autoref{fig:thermoforming} illustrates the whole process of transferring a 3D model into a plastic replica with the original texture applied atop it.

\subsubsection{Hybrid Fabrication Techniques}
\label{HybridFab}

In our survey, we refer to the fabrication method of a work as \textit{hybrid} when a series of various methods have been used to make one single physicalization. In other words, if a physicalization system produces different results, each with one single fabrication technique, it will not be counted as a hybrid method in our work. 

There are \numHybridFab examples of hybrid fabrication methods in our corpus. In some cases, hybrid approaches have been taken to deal with the issue of reproducing colour in physicalizations by using inkjet printing and 3D printing (e.g.,\cite{djavaherpour2017physical,moorman2020geospatial,oh2018pep}) or inkjet printing and CNC milling (e.g.,\cite{Solid2003Airplane}). In other examples, some parts of a physicalization are made with one tool and some parts through another method, based on the specs and limitations of each fabrication technique. Examples of such hybrid works are laser cutting (or waterjet cutting) and 3D printing (e.g.,\cite{katsumoto2018robotype,Kang2017Wearable,Allen2016Motus}), CNC and 3D printing (e.g.,\cite{Shumay2016FizViz}), CNC and laser cutting (e.g.,\cite{yao2015biologic}), and 3D printing, laser cutting, and digital embroidery for making personalized shoes\cite{nachtigall2019encoding}.

\subsubsection{Manual Assembly}
\label{subsubsec:assembly}

There are \numAssembly works in our corpus that are fabricated in separate pieces and need to be assembled to form the final physicalization. Digital fabrication machines have a limited build area (for additive tools) and support specific sizes for sheets and blocks of material (for subtractive and cutting tools). This limitation forces the design to be either limited to a scale that can be fabricated in one piece or to be piece-wise in a way that can be assembled and make a bigger scale physicalization (e.g.,\cite{allahverdi2018landscaper,marcus2014centennial,Huang2017moire}. Assembling a physicalization also provides various interaction opportunities (e.g.,\cite{gourlet2017cairn}) and can be used for educational purposes (e.g.,\cite{nadeau2000visualizing,moorman2020geospatial}) (see Section \ref{sec:discussion} for more details). 



\subsection{Augmenting Physicalizations}
\label{subsec:augment}

As introduced in Section \ref{subsec:passive}, augmenting a physicalization adds an extra layer of information to an otherwise passive physical object and can be a straightforward way of adding sophisticated interactivity without integrating actuators (e.g., motors) as is required for active physicalizations. The augmentation approaches observed in our corpus can be summarized as a form of augmented reality, realized by projecting directly onto the physical object or through a personal AR view using \ a head-mounted display or a hand-held device equipped with a camera. Overall relatively few of the items in our corpus (\numAugmented out of \numTotal) fall into the category of augmented physicalizations.


\subsubsection{Projection Augmentation} 
Physicalizations using projection augmentation consist typically of a passive, fabricated physicalization with additional data projected directly onto it. In some cases, projections provide additional data layers, such as annotations~\cite{priestnall2012projection}. In other cases, they permit interactivity, such as highlighting~\cite{hemment2013emoto}. Examples in our corpus include relief maps (\cite{tateosian2010tangeoms,priestnall2012projection,millar2018tangible} (see \autoref{fig:geospatial}), globe-based time-varying geospatial data \cite{dadkhahfard2018area} and a data sculpture (\cite{hemment2013emoto}) that shows Twitter sentiment data as an abstract relief heatmaps.

When using projection augmentation, it is necessary to calibrate the physical object and the projection so that projected information lines up with corresponding physical features. The TanGeoMS system~\cite{tateosian2010tangeoms} includes a combination of projection and 3D-scanning which enables the system to recognize the topology of the passive physical model and automatically detect how to rotate and scale the topological data to be projected onto it. For the PARM~\cite{priestnall2012projection} and Emoto systems~\cite{hemment2013emoto}, no calibration details are provided;  most likely, they require manual calibration between the physical model and the projection. When projecting on non-flat surfaces, it is often necessary to apply some form of projection mapping~\cite{grundhofer2018recent} to avoid visible distortions of the projected content. This is only discussed for the TanGeoMS system, where they found that a correction would only be necessary for height differences of more than 6 cm, which did not occur in their case. The other two projection-augmented physicalizations did not discuss applying any remapping; this may be due to the small height differences present in the two examples.




\subsubsection{Augmentation through Personal Augmented Reality (AR)}
Unlike projected augmentation, personal augmented reality (AR) are technologies (head-mounted displays, mobile devices) that offer an augmented perspective of an object to a single individual. Augmentation through personal augmented reality is less present in our corpus (only \numAR items out of \numTotal examples make use of an individualized augmented reality review). Of these, two are academic works that date from before the consumer-availability of augmented reality headsets (2004/2005); these projects display the augmentation layer on a separate display overlaid on a live video feed of the physicalization~\cite{gillet2004computer, gillet2005tangible}. The third example, PLANWELL~\cite{nittala2015planwell}, uses tablets and mobile phones to display augmentations. While this has the advantage of combining camera and display and aligning their viewpoint, it also becomes more cumbersome to use: these devices must be held which both limits the use of the users' hands and occludes their view of the actual physicalization. 

While our corpus does not include any examples using head-mounted displays (HMD), this is an obvious and promising avenue to explore. However, in such cases, every user needs to be equipped with an HMD to be able to view and benefit from the augmentation. This is not the case with systems using projection where anyone in their proximity is able to view the augmentation. These tradeoffs are not specific to physicalizations and have already been discussed elsewhere, for example, by Thomas et al.~\cite{thomas2018situated}.


\begin{figure*}[t]
  \centering
   \includegraphics[bb=0 0 8691 1706, width=\textwidth]{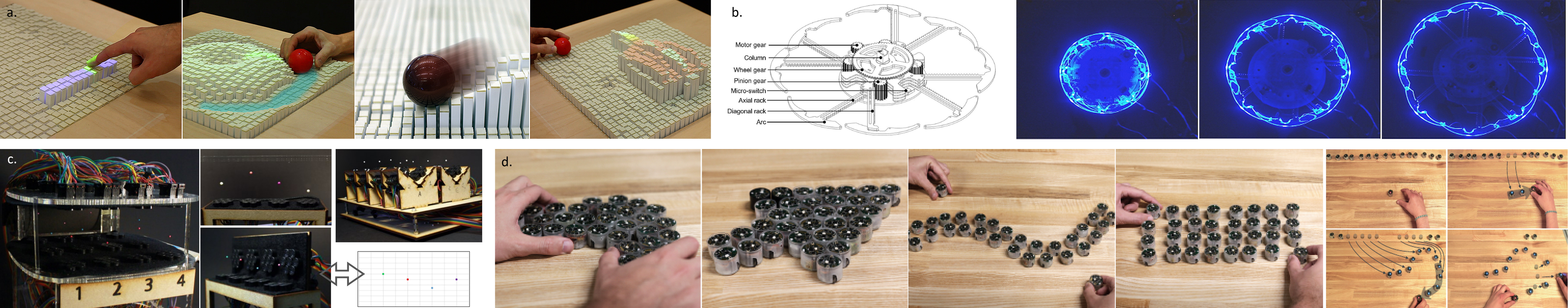}
   \caption{Active physicalizations use different techniques to support physical changes: (a) inFORM is a shape-changing display that enables several new interaction techniques. Images taken from\cite{follmer2013inform}, (b) Expandable and stackable actuating rings can show different datasets. Images taken from\cite{daniel2018designing}, (c) Ultrasound can be used to control the movement of lightweight objects within the range of actuators. Images taken from\cite{omirou2016floating}, (d) Zooids proposed the use of small moving robots to represent several scenarios of data and interactions. Images taken from\cite{le2016zooids}.}
           \label{fig:active}
\end{figure*}

\subsection{Active Physicalization Rendering Techniques}
\label{subsec:active}

Active physicalization rendering techniques go beyond what is possible through augmentation, but they are generally also more difficult to realize. Nonetheless, they are more represented in our corpus with \numActive examples. Many of these examples fall into the category of data sculptures where artists explored unique ways to actuate materials in some way suited to communicate their artistic intent or where academic authors sought to find unique ways of representing often personal data in appealing ways. We focus here on examples that are not entirely specific to the context in which they were created and whose analysis can inform the design of future active physicalizations in some way.   

Overall, we identify three main goals for choosing an active rendering technique: supporting changes of a single dataset, supporting multiple datasets, and enabling interactivity beyond what is possible using augmentation approaches. These goals are often combined although interactivity is less common for data sculptures/installations. Orthogonal to these goals is the question where the rendering technique should be capable of dealing with varying numbers of data points or whether those remain fixed once chosen.



If an active physicalization only supports changes to a single dataset, then this suggests that the rendering technique was probably specifically developed or tailored to that dataset and may only be applied to other data with difficulty. This is something we mostly observed with examples classified as data sculptures. For example, the artist Rafael Lozano-Hemmer developed multiple installations falling into this category using situated data like the presence and location of people in a room to actuate belts~\cite{Lozano2004Array} or tape measures~\cite{Lozano2011Tape}. The latter example (shown in \autoref{fig:statistical}d) uses actuated tape measures going up and down which resembles a bar chart and thus could also be used with different datasets. The number of data points of such a system remains fixed though and would need to be manually extended to be usable with different data sets. Both of these examples react to people's presence in the room, that is, sensors capture their presence and reflect the data on the shape and orientation of the system. Beyond that, these installations offer no interactivity and they are purely meant to present data and not to enable onlookers to explore the shown data in any way. 

A few platforms have been proposed, mostly in academic research, which enable the visualization of various datasets as well as interactivity aspiring to achieve a level of functionality known from web-based visualization tools, such as, support to view different data, searching, filtering, highlighting etc. The development of such platforms generally requires skills in mechanics, fabrication, sensor and actuator choice and placement, and micro-controller development. Reviewing all the issues related to developing new active platforms would go beyond the scope of this article. We review here only active platforms included in our corpus. 




\subsubsection{Shape-changing Displays}
Shape-changing displays are actuated devices capable of deforming in various ways~\cite{rasmussen2012shape}. One item in our corpus, the Xenovision III system, is an actuated solid terrain model, commercially available and marketed for military applications~\cite{Schmitz2004Xenovision}. It is capable of displaying any terrain data using its 7,000 actuated pins. Most of the other shape-changing displays in our corpus are created either by academics as proofs-of-concepts or by artists for installations in museums or galleries. The most common form factor for such displays are rods or bars arranged in arrays and capable of moving up and down to provide a 2.5D display~\cite{taher2015exploring,follmer2013inform} (see \autoref{fig:statistical}a/b and \autoref{fig:active}a). Such displays have generally much fewer actuators than the Xenovision system, that is between 100 and 1,000 pins. All of these 2.5D systems cannot display any overhangs and only show data that could be represented by a 3D barchart resulting often in a resolution of 5 to 10 mm per actuator. The Relief prototype uses a similar principle but connected the individual actuators with a cloth such that a smoothed surface is created~\cite{leithinger2011direct} which lends itself naturally to display terrain data. Shape-changing displays can also come in different base shapes. For example, Daniel and colleagues used a ring shape and actuated their display such that rings could be stacked and each ring could expand its size to show different data~\cite{daniel2018designing} (see \autoref{fig:active}b). All of these displays can generally show different data sets or update the data being shown currently. Most also support interactivity in some way, often by covering the interactive area with a depth camera and subsequently interpreting people's gestures around the devices.


\subsubsection{Suspended elements} 
While two-dimensional visualizations on-screen or three-dimensional visualizations in virtual reality are free to render data points where the data demands them to be shown, physicalizations are bound by the laws of physics. There have been, however, a few attempts to overcome these limitations. A common approach, especially with data sculptures, is the use of strings or ropes to suspend and actuate sail-like structures~\cite{Keller2009DataMorphose} or spheres~\cite{BMW2008Kinetic, Leng2012Point}. Few have attempted to suspend elements without any support: Omirou and colleagues made use of ultra-sound to control the position and movement of small and lightweight objects within the range of their actuators~\cite{omirou2016floating} (see \autoref{fig:active}c). While this is a promising direction to realize physical 3D scatterplots, the authors report limitations to what data can be shown since particles can disturb each other when placed in close proximity. A possibly different approach is the use of magnetic forces as illustrated by Lee and colleagues~\cite{lee2011zeron}. Note that this work is not part of our corpus since the authors only illustrated the levitation of one element which would only permit the physicalization of very trivial data. Physicalizations with suspended elements tend to be less interactive than those using shape displays. This may be due to user interaction potentially interfering with the technology used to suspend elements.

\subsubsection{Robotic approaches} 
A few examples in our corpus have used robotic arms to assemble physicalizations. While the overall rendering platform can be considered active -- they take data as input and render a physical object -- these resulting objects are, once assembled, passive objects. Our corpus includes three examples falling into this category. One uses a Kuka industrial robot to place nails in a substrate according to wind data~\cite{Aweida2013RobotArranges}, one uses a similar type of robot to span strings to approximate the visual shape of the input data~\cite{birsak2018string}, and a third uses a self-built robotic system based on robotic toys and vacuum cleaners to select already printed paper pie charts and place them on actual pies~\cite{rust2014piece}. 

An entirely different approach was taken by Le Goc and colleagues who developed a platform of small robots they called zooids~\cite{le2016zooids} (see \autoref{fig:active}d). Each of these robots is meant to represent one data point and to move around a surface covered by structured light within which it can orient itself and move to show different facets of the data point it represents~\cite{LeGoc2018dynamic}. While the examples using robotic arms are active platforms producing passive physicalizations of different data sets, zooids is a truly active physicalization system capable of showing many different data sets and types and allow people to interact with the robots by picking them up or moving them around or react to system commands asking them to sort or rearrange the data.









\subsection{Automated Physical Rendering Platforms}
\label{subsec:automated}

\begin{figure*}[t]
  \centering
   \includegraphics[bb=0 0 2443 484, width=\textwidth]{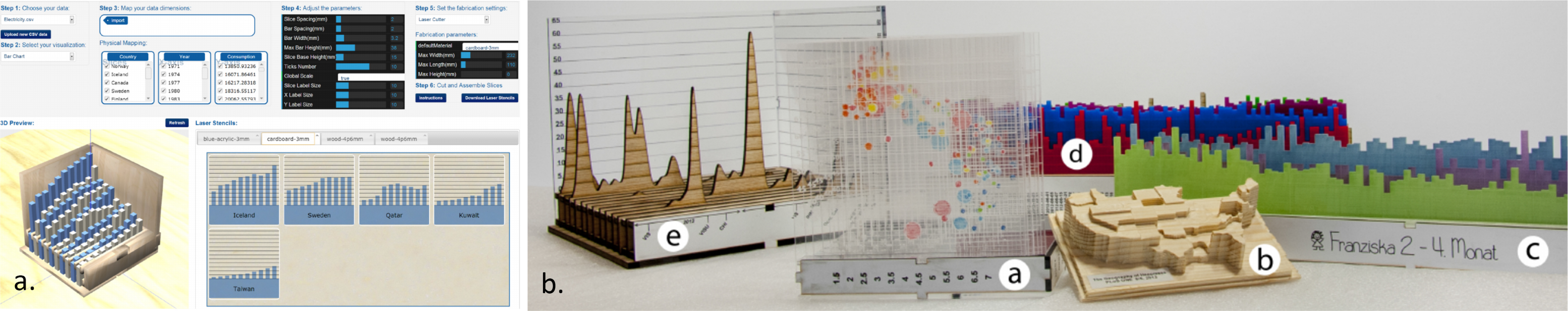}
   \caption{MakerVis is an automated software developed for making physicalizations. (a) A screen-capture of the UI, (b) Physicalizations made by MakerVis. Images taken from \cite{swaminathan2014supporting}.}
           \label{fig:MakerVis}
\end{figure*}

The rendering process of a physicalization work, from the early design stages to fabrication and assembly, is a skill-oriented approach. In other words, it demands knowledge and expertise in data visualization, digital design, and digital fabrication, and is sometimes involved with labour-intensive craftsmanship based on the applied fabrication techniques. In order to overcome this issue, some research has been undertaken with the goal of automating the whole or parts of the physical rendering process. \textit{MakerVis} \cite{swaminathan2014supporting} is one of the most inclusive platforms developed for this goal that is capable of automating the whole physical rendering process from data filtering to physical fabrication. The prototype software of MakerVis reads data in CSV or topoJSON (for prism maps) formats and can produce data types that are compatible with CNC machining, 3D printing, and laser cutting. The software is a web application built on top of NodeJS, D3, JQuery, and ThreeJS frameworks. Figure \ref{fig:MakerVis} shows the interface of MakerVis, as well as some results made by it.

\section{Discussion}
\label{sec:discussion}

In this section, we discuss some of the decisions and challenges of transforming a data physicalization concept into a fabricated physical form.  First, we briefly discuss general digital fabrication issues that remain a challenge when rendering data physicalizations: design for manufacture and assembly, and prototyping and iterative design. Then, we discuss challenges that impact both fabrication and data representation. When consulting with stakeholders, data physicalization designers are trying to understand both the physical and representational requirements of the final object. When deciding on scale, the designer must balance scale limitations in manufacturing against the readability or user experience of the physicalization. Finally, assembly decisions can also limit users' ability to interact with data in the final object.





\subsection{Common Digital Design and Fabrication Challenges}

Many of the physicalizations from our sample struggled with challenges that are universal to all digitally fabricated forms.

\subsubsection{Design for Manufacture and Assembly}

When optimizing a design for fabrication, designers often adjust the features or complexity of a digital design to reduce manufacturing time (e.g., \cite{khot2014understanding}). Many of the physicalizations in our sample required cutting and stacking layers of plastic or wood (e.g., \cite{hemment2013emoto,Hush2014MadebyNum,Stusak2013Layered,Adrien2011data}).  However, depending on the geometric complexity of each layer and the number of layers, this approach can be time-consuming for both fabrication and assembly of the final structure. We found two examples \cite{djavaherpour2017physical, marcus2014centennial} of physicalizations that included instructions to expedite the final assembly.

\subsubsection{Prototyping and Iterative Design}

To facilitate iterative design, physicalization designers used both digital simulation and lower-fidelity prototyping techniques to validate designs before following-through with the final version.  Prototyping helps designers make better decisions about the specifications of the physicalization; effective prototyping techniques require the least amount of material or time investment to obtain insights to drive the next series of design changes.

In a digital design process, digital simulations of the fabrication process or the final object can facilitate faster iterations on a physicalization design. This includes active physicalizations, where Kangaroo Plug-in for Rhino Grasshopper can be used to simulate movements of active structures (e.g.,  \cite{velikov2014nervous,Aweida2013RobotArranges,thun2012soundspheres}). 

Our sample also included many examples of physical prototypes~\cite{ang2019physicalizing,barrass2011phsyical,marcus2014centennial}, which offer quicker, lower-cost interim representations for design iteration. 
One approach is to build a small portion of the final physicalization, with the final scale, fabrication technique, and material as done by~\cite{allahverdi2018landscaper}. 
Another approach is to build a scale-model of a larger-scale physicalization as a way of verifying the overall design and rationale (e.g., \cite{hosseini2019data}). 
Another approach is to prototype using a lower-fidelity fabrication technique. For example, in bioLogic\cite{yao2015biologic}, researchers developed a process of compositing conductive traces to the existing biofilm via screenprinting; as a lower-fidelity alternative, they also laser cut double-sided tacky paper to create masks for screenprinting.

\subsection{Collisions between Data Representation and Digital Fabrication}

In the following section, we discuss three areas where digital fabrication challenges and data representation challenges collide -- understanding users' requirements; physical scale; assembly and interaction.


\subsubsection{Understanding Users' Requirements}

Some physicalization designers conducted formative user research to guide what data needed to be represented by the physicalization, as well as any physical requirements of the physicalization object. This early decision-making process may include interviews or consultations with stakeholders' about their own understanding of their data or expectations on the form of a resulting physicalization ~\cite{khot2014understanding,van2018exploring}.


Consultation with end-users can inform which representational idioms are most appropriate for the target community.  Many papers in our example approached specific user groups to validate their choice of physicalization idiom \cite{priestnall2012projection,gwilt2012enhancing,khot2014understanding,gillet2005tangible,priestnall2012projection,keefe2018weather,lee2015patina}. For example, Gwilt et al.\cite{gwilt2012enhancing} found that an engineering community preferred bar and pie charts, whereas a design community preferred data sculptures. Similarly, Khot et al.\cite{khot2014understanding} found their community of exercise enthusiasts preferred to represent their physical activity using a non-scientific idiom (size of a frog) over a more scientific representation (physical bar chart).
Meanwhile, molecular biologists from Gillett et al.\cite{gillet2005tangible} preferred augmented physicalizations over static physicalizations.

\subsection{Physical Scale}

The physical scale of a physicalization fundamentally changes how people interact with it as an object; it also can introduce pragmatic constraints such as weight, balance, or portability of the final object. 
In room-size physicalization installations or spatializations, such as~\cite{Lozano2004Array,Lozano2011Tape,Barry2012Saturation,BMW2008Kinetic}) and  \cite{marcus2014centennial,velikov2014nervous,hosseini2019data,keefe2018weather,Huang2017moire}, viewers must move around the space defined by the physicalization to explore various aspects of the data. \
Meanwhile, table-top physicalizations~\cite{tateosian2010tangeoms,allahverdi2018landscaper}, including augmented physicalizations~\cite{dadkhahfard2018area,millar2018tangible,kirshenbaum2020data}, still require viewers to move around them, but there is no specific space created by them that helps with the exploration of data. 
Hand-held physicalizations (e.g., \cite{khot2014understanding,kane2014tracking,ang2019physicalizing,bader2018making}) provide very different interaction scenarios, as users can easily manipulate them; objects that must be held in-hand (e.g., \cite{patel20173d,tymms2018quantitative} can leverage tactile cues (surface roughness or texture). 

However, the scale of a physicalization also majorly influences which fabrication techniques are appropriate. Most fabrication techniques work at specific scales; for exmaple, most FDM 3D printers have an average build volume of around $20^3 cm^3$. 
Larger-scale visualizations can be fabricated modular components at the scale of fabrication machinery, and then assembling these components into a larger structure. Decomposing a large 3D object into smaller parts that fit in the printing volume was introduced by Chopper\cite{luo2012chopper} and then used by several physicalization works (e.g.,\cite{allahverdi2018landscaper,dadkhahfard2018area,djavaherpour2017physical,fetterman2014luminocity}). 
However, these techniques rely more heavily on computationally-generated instructions for manual assembly. For instance, it was impossible for high school students and their geography teachers in Australia to assemble the modular globes of eastern Australia and western Canada without any indexing, due to the unique geometry of the pieces\cite{moorman2020geospatial}.
Meanwhile, small-scale visualizations are constrained by the resolution of the fabrication technique. For example, within the realm of 3D printing, Fuse-Deposition Modeling (FDM) 3D printers cannot make features smaller than its extrusion head; Stereolithography (SLA) printers can make features as small as the laser used to cure resin (e.g.,~\cite{Allen2016Motus,zhang2018mosculp}. 

\subsection{Assembly and Interaction}
\label{subsec:assembly-interaction}

As with any digital design, when physicalizations are broken into several pieces for fabrication purposes, the designer must define how those pieces connect to each other.  This includes defining which attachment techniques will be used, specific feature parameters to assembly features (e.g., joint location, feature dimensions, part clearance between parts), as well as the assembly process itself.

However, the way in which a physicalization is designed to be assembled can determine what types of interactions end-users can have with its represented data. For example, Jansen et al.\cite{jansen2013evaluating} created modular 3D bar charts that allowed the end-user to select, reorder, and independently compare datasets. If instead these bar charts had been glued together, the end-user would no longer be able to interact as deeply with the data itself.  Similarly, the act of assembling a dataset can require end-users to more deeply interact with the data itself.  This is a guiding principle of both constructive visualization\cite{huron2014constructive} and participatory visualization (e.g., \cite{gourlet2017cairn}), but is also present in many other physicalizations in our sample~\cite{khot2016fantibles,allahverdi2018landscaper,Schneider2012Lego,djavaherpour2017physical,moorman2020geospatial}.  Landscaper\cite{allahverdi2018landscaper} requires assembling a series of geospatial features (e.g., trees, rocks, road networks, and urban features) which simultaneously requires the end-users to become familiar with where each feature belongs within the space. 
In Nadeau et al~\cite{nadeau2000visualizing}, end-users could detach the 1:1 interlocking scale model of the skull and brain segmentation to better understand the complex volumetric dataset via cutting operations; this cross-section is only possible when this type of disassembly is pre-planned and allowed. Vol$^2$velle\cite{stoppel2016vol} introduced a novel interaction system that physically recreates the traditional concept of Volvelle.

\section{Conclusion and Future Work}
\label{sec:conclusion}

Physicalizations are effective tools for conveying the message of various datasets and they can be rendered in many different methods, for various applications, and in many representational idioms. Although many impressive works have been done in this field, many areas of rendering physicalizations remain unexplored. For instance, there have been many creative fabrication methods introduced by the computer graphics community with great potentials for physicalizations, especially for dealing with the issue of colour and texture. By applying such methods to physicalization rendering, many interesting possibilities will be introduced. Also, numbers in our corpus show that physicalization has been slightly underexposed in the scientific community and for pragmatic purposes. Many interesting aspects of physical rendering in artistic physicalizations and interactive installations have strong potentials to be applied to scientific works as well. In our survey, we discussed that augmented physicalizations add extra layers of information to passive works and make them more sophisticated objects to interact with. However, very few efforts have been made to develop such works. With the increasing popularity and technological advances of devices that support augmented reality, physicalization designers should take advantage of elevating their passive designs to the next level of informative and interactive representations. 

In this survey, we have provided an overview of physicalizations, their classifications, visual representation formats, and their target datasets. More importantly, we have reviewed various methods that physicalizations can be designed in digital design approaches and then rendered physically by digital fabrication tools. We hope that computer graphics, visualization, interaction, art, industrial design, and architecture communities find this survey useful and a source of inspiration to develop the physicalization field further.

\newpage

\newpage
\setlength\tabcolsep{2.5pt} 

\newcolumntype{R}[2]{%
    >{\adjustbox{angle=#1,lap=\width-(#2)}\bgroup}%
    l%
    <{\egroup}%
}

\begin{table*}[htp!]
\centering
\resizebox{.83\textwidth}{!}{%
\begin{tabular}{ c | c | c | c | ccc | ccccccc | ccccc | ccccccc | ccccc | cc }
\toprule
\multicolumn{4}{c|}{Basic Info} &\multicolumn{3}{c|}{Class} &\multicolumn{7}{c|}{Dataset(s)} &\multicolumn{5}{c|}{Category} &\multicolumn{7}{c|}{Digital Fabrication} &\multicolumn{5}{c|}{Digital Design} &\multicolumn{2}{c}{} \\ 
\midrule
\cellcolor[HTML]{F6F6F6} \rotatebox{90}{Provenance} &\cellcolor[HTML]{F6F6F6}\rotatebox{90}{Year} &\cellcolor[HTML]{F6F6F6}\rotatebox{90}{Title} &\cellcolor[HTML]{F6F6F6}\rotatebox{90}{Type} &\cellcolor[HTML]{F6F6F6}\rotatebox{90}{Passive} &\cellcolor[HTML]{F6F6F6}\rotatebox{90}{Active} &\cellcolor[HTML]{F6F6F6}\rotatebox{90}{Augmented } &\cellcolor[HTML]{F6F6F6}\rotatebox{90}{Bio/Medical} &\cellcolor[HTML]{F6F6F6}\rotatebox{90}{Math} &\cellcolor[HTML]{F6F6F6}\rotatebox{90}{Personal} &\cellcolor[HTML]{F6F6F6}\rotatebox{90}{Geospatial} &\cellcolor[HTML]{F6F6F6}\rotatebox{90}{Image/Video} &\cellcolor[HTML]{F6F6F6}\rotatebox{90}{Statistical} &\cellcolor[HTML]{F6F6F6}\rotatebox{90}{Other} &\cellcolor[HTML]{F6F6F6}\rotatebox{90}{InfoVis} &\cellcolor[HTML]{F6F6F6}\rotatebox{90}{SciVis} &\cellcolor[HTML]{F6F6F6}\rotatebox{90}{Pragmatic} &\cellcolor[HTML]{F6F6F6}\rotatebox{90}{Artistic} &\cellcolor[HTML]{F6F6F6}\rotatebox{90}{Data Sculpture} &\cellcolor[HTML]{F6F6F6}\rotatebox{90}{3D Print} &\cellcolor[HTML]{F6F6F6}\rotatebox{90}{CNC} &\cellcolor[HTML]{F6F6F6}\rotatebox{90}{Laser Cut} &\cellcolor[HTML]{F6F6F6}\rotatebox{90}{Robotic Arm} &\cellcolor[HTML]{F6F6F6}\rotatebox{90}{Hybrid} &\cellcolor[HTML]{F6F6F6}\rotatebox{90}{Other} &\cellcolor[HTML]{F6F6F6}\rotatebox{90}{Assembly?} &\cellcolor[HTML]{F6F6F6}\rotatebox{90}{CAD} &\cellcolor[HTML]{F6F6F6}\rotatebox{90}{Parametric} &\cellcolor[HTML]{F6F6F6}\rotatebox{90}{Programming} &\cellcolor[HTML]{F6F6F6}\rotatebox{90}{Hybrid} &\cellcolor[HTML]{F6F6F6}\rotatebox{90}{Other} &\cellcolor[HTML]{F6F6F6}\rotatebox{90}{Iterative Design} &\cellcolor[HTML]{F6F6F6}\rotatebox{90}{User Study} \\
\bottomrule
\end{tabular}%
}
\caption{Taxonomy of the works reviewed in this survey (Part 1).} \label{tab:Taxonomy1} 
\end{table*}

\newpage
\setlength\tabcolsep{2.5pt} 



\begin{table*}[!htp]
\centering
\resizebox{.83\textwidth}{!}{%
\small
\begin{tabular}{ c | c | c | c | ccc | ccccccc | ccccc | ccccccc | ccccc | cc }
\toprule
\multicolumn{4}{c|}{Basic Info} &\multicolumn{3}{c|}{Class} &\multicolumn{7}{c|}{Dataset(s)} &\multicolumn{5}{c|}{Category} &\multicolumn{7}{c|}{Digital Fabrication} &\multicolumn{5}{c|}{Digital Design} &\multicolumn{2}{c}{} \\
\midrule
\cellcolor[HTML]{F6F6F6}\rotatebox{90}{Provenance} &\cellcolor[HTML]{F6F6F6}\rotatebox{90}{Year} &\cellcolor[HTML]{F6F6F6}\rotatebox{90}{Title} &\cellcolor[HTML]{F6F6F6}\rotatebox{90}{Type} &\cellcolor[HTML]{F6F6F6}\rotatebox{90}{Passive} &\cellcolor[HTML]{F6F6F6}\rotatebox{90}{Active} &\cellcolor[HTML]{F6F6F6}\rotatebox{90}{Augmented } &\cellcolor[HTML]{F6F6F6}\rotatebox{90}{Bio/Medical} &\cellcolor[HTML]{F6F6F6}\rotatebox{90}{Math} &\cellcolor[HTML]{F6F6F6}\rotatebox{90}{Personal} &\cellcolor[HTML]{F6F6F6}\rotatebox{90}{Geospatial} &\cellcolor[HTML]{F6F6F6}\rotatebox{90}{Image/Video} &\cellcolor[HTML]{F6F6F6}\rotatebox{90}{Statistical} &\cellcolor[HTML]{F6F6F6}\rotatebox{90}{Other} &\cellcolor[HTML]{F6F6F6}\rotatebox{90}{InfoVis} &\cellcolor[HTML]{F6F6F6}\rotatebox{90}{SciVis} &\cellcolor[HTML]{F6F6F6}\rotatebox{90}{Pragmatic} &\cellcolor[HTML]{F6F6F6}\rotatebox{90}{Artistic} &\cellcolor[HTML]{F6F6F6}\rotatebox{90}{Data Sculpture} &\cellcolor[HTML]{F6F6F6}\rotatebox{90}{3D Print} &\cellcolor[HTML]{F6F6F6}\rotatebox{90}{CNC} &\cellcolor[HTML]{F6F6F6}\rotatebox{90}{Laser Cut} &\cellcolor[HTML]{F6F6F6}\rotatebox{90}{Robotic Arm} &\cellcolor[HTML]{F6F6F6}\rotatebox{90}{Hybrid} &\cellcolor[HTML]{F6F6F6}\rotatebox{90}{Other} &\cellcolor[HTML]{F6F6F6}\rotatebox{90}{Assembly?} &\cellcolor[HTML]{F6F6F6}\rotatebox{90}{CAD} &\cellcolor[HTML]{F6F6F6}\rotatebox{90}{Parametric} &\cellcolor[HTML]{F6F6F6}\rotatebox{90}{Programming} &\cellcolor[HTML]{F6F6F6}\rotatebox{90}{Hybrid} &\cellcolor[HTML]{F6F6F6}\rotatebox{90}{Other} &\cellcolor[HTML]{F6F6F6}\rotatebox{90}{Iterative Design} &\cellcolor[HTML]{F6F6F6}\rotatebox{90}{User Study} \\
\bottomrule
\end{tabular} }
\caption{Taxonomy of the works reviewed in this survey (Part 2).}
\label{tab:Taxonomy2}
\end{table*}

\clearpage
\newpage
\clearpage
\newpage

\section{Authors' Short Biographies}
\label{sec:bio}

\textbf{Hessam Djavaherpour} is a Ph.D. Candidate in Computational Media Design (CMD) at the University of Calgary. He is an architect and a computational designer interested in physical visualization of data, data-driven design approaches, algorithmic and parametric design, and digital fabrication. His core Ph.D. publications are mainly focused on pursuing the concept of physicalizing geospatial data at various scales and applications, such as physicalization of a partial globe~\cite{djavaherpour2017physical}, 3D printing landscapes~\cite{allahverdi2018landscaper}, geospatial physyicalization as an urban structure~\cite{hosseini2019data}, and studying the efficiency of physicalization in geography education~\cite{moorman2020geospatial}.

\noindent\textbf{Faramarz Samavati} is a professor of Computer Science at the University of Calgary. Dr. Samavati's research interests include
Computer Graphics, Visualization, and Digital Earth. Over the past eight years, he has received seven best paper awards, Digital Alberta Award, Great Supervisor Award, University of Calgary Peak Award and Faculty of Science Established Career Scholarship Excellence Award. He has supervised several graduate students with theses related to physicalization. He has published many papers in this area, including geospatial physicalization~\cite{djavaherpour2017physical, allahverdi2018landscaper, moorman2020geospatial}, physicalizing cardiac blood flow data via 3D printing~\cite{ang2019physicalizing}, urban structure and heritage~\cite{hosseini2019data}, and a system for covering fabricated objects~\cite{mahdavi2015coverit}.  Currently, in addition to continuing his research on geospatial physicalization, he works on the physicalization of temporal bones for surgery simulation and multiscale physicalization of historical sites.

\noindent\textbf{Ali Mahdavi-Amiri} is currently a University Research Associate in the Department of Computing Science at Simon Fraser University.
His primary research interest is in visual computing with focus on geometry processing, computational fabrication, and machine learning.
His work on computational fabrication covers a variety of research problems and fabrication techniques such as machining optimization (3-axis CNC)\cite{mahdavi2020VDAC}, fabricating assembly puzzles and toys (3D printers)\cite{mahdavi2015coverit,li2018construction}, geospatial physicalization (3D printers)\cite{djavaherpour2017physical,allahverdi2018landscaper}, and illusive structures (laser cutters)\cite{hosseini2020portal}.

\noindent\textbf{Fatemeh Yazdanbakhsh} is a Ph.D. student at the University of Calgary working on the application of physicalization for producing medical prototypes. Her research focus is on making physical models as a replacement for cadaveric bones used for temporal bone surgery rehearsal and teaching purposes. She explores different materials to find the best match for reproducing tactile sense for hard and soft tissue. Her research also includes finding a method to fabricate complicated structures in different colours and materials using off-the-shelf 3D printers.

\noindent\textbf{Samuel Huron} is an associate professor in Design of Information Technologies inside the Social and Economical Science Department of Telecom Paris School at the Institut Polytechnique de Paris, and part of the CNRS Institut Interdisciplinaire of innovation. He is deeply interested in how the construction process of physical representation of data impact the cognitive process of the authors. He has worked extensively on the topic of constructing data physicalisation, as a design paradigm~\cite{huron2014constructive}, as a way to study and understand visual mapping process~\cite{huron2014constructing}, as a way to facilitate self reflexion~\cite{thudt2018self}, as a way to compare authoring process with digital tools~\cite{wun2016comparing}. Since the last 7 years, he also organized many workshops in academic settings such as IEEE VIS~\cite{hogan2018}, ACM DIS~\cite{hogan2017pedagogy}, DRS~\cite{huron2017let} and also in a multitude of non-academic setting.

\noindent\textbf{Richard Levy} has recently retired from the University of Calgary where he was a Professor of Planning and Urban Design at The University of Calgary for 26 years. Dr. Levy also served as the Co-Director of the Computational Media Program (CMD) and is an Adjunct Professor in the Department of Computer Science and The Department of Archaeology. Dr. Levy has conducted research projects with faculty from  Archaeology, Computer Science, Geomatics Engineering, Kinesiology and Psychology. Dr. Levy speaks at international and national conferences in the fields of archaeology, education, serious games, urban planning, and virtual reality.

\noindent\textbf{Yvonne Jansen} is a tenured research scientist with the French National Center for Scientific Research (CNRS) and a member of the Institute for Intelligent Systems and Robotics at Sorbonne Université. She has published extensively on the topic of data physicalization including a handbook chapter \cite{dragicevic_data_2021}, a review and research agenda \cite{jansen2015opportunities}, an interaction model \cite{jansen2013interaction}, authoring tools with Lora Oehlberg \cite{swaminathan2014supporting}, and empirical work of both quantitative \cite{jansen2013evaluating} as well as qualitative nature with Samuel Huron \cite{huron2014constructing}. She also organized a Dagstuhl seminar on data physicalization as well as workshops at multiple conferences including CHI and VIS. Together with Pierre Dragicevic she curates the list of physical visualizations and related artifacts (\url{http://dataphys.org/list/}).

\noindent\textbf{Lora Oehlberg} is an Associate Professor of Computer Science at the University of Calgary. Her background is in design theory in methodology, which she applies to the design of technologies to support creativity and collaboration. She has published several papers at the intersection of physical authoring tools and data visualization, most notably using physical graphical template tools for \cite{wun2019you}, considering alternative fabrication media for data physicalization \cite{wannamaker2019data} and proposing a physicalization authoring tool with Yvonne Jansen \cite{swaminathan2014supporting}. She has also participated in workshops at VIS 2019~\cite{Oehlberg2018encoding} and at Dagstuhl considering new frontiers of data physicalization.

\clearpage
\newpage

\clearpage
\newpage

\bibliographystyle{eg-alpha-doi}  
\bibliography{bibliography}        

 

\end{document}